\documentclass[draftclsnofoot]{IEEEtran}
\usepackage{graphicx}
\usepackage{epstopdf}

\usepackage{subfigure}
\usepackage{booktabs}
\usepackage{caption}
\usepackage{color}
\usepackage{bm}
\usepackage{CJK}
\usepackage{indentfirst}
\usepackage{amsmath}
\usepackage{amssymb}
\usepackage{float}
\usepackage{multirow}
\usepackage{algorithm}
\usepackage{algpseudocode}
\usepackage{amsmath}
\usepackage{flushend}
\usepackage{geometry}
\newtheorem{theorem}{Theorem}

\begin{document}
\title{Multi-timescale Channel Customization for Transmission Design in RIS-assisted MIMO Systems}
\author{\IEEEauthorblockN{Weicong~Chen, Chao-Kai~Wen, Xiao~Li, and~Shi~Jin}\\
\thanks{{Weicong Chen, Xiao Li, and Shi Jin are with the National Mobile Communications
Research Laboratory, Southeast University, Nanjing 210096, China (e-mail: cwc@seu.edu.cn; li\_xiao@seu.edu.cn; jinshi@seu.edu.cn).} }
\thanks{Chao-Kai Wen is with the Institute of Communications Engineering,
National Sun Yat-sen University, Kaohsiung City 80424, Taiwan (e-mail:
chaokai.wen@mail.nsysu.edu.tw).}
}

\maketitle
\begin{abstract}
 The performance of transmission schemes is heavily influenced by the wireless channel, which is typically considered an uncontrollable factor. However, the introduction of reconfigurable intelligent surfaces (RISs) to wireless communications enables the customization of a preferred channel for adopted transmissions by reshaping electromagnetic waves. In this study, we propose multi-timescale channel customization for RIS-assisted multiple-input multiple-output systems to facilitate transmission design. Specifically, we customize a high-rank channel for spatial multiplexing (SM) transmission and a highly correlated rank-1 channel for beamforming (BF) transmission by designing the phase shifters of the RIS with statistical channel state information in the angle-coherent time to improve spectral efficiency (SE). We derive closed-form expressions for the approximation and upper bound of the ergodic SE and compare them to investigate the relative SE performance of SM and BF transmissions. In terms of reliability enhancement, we customize a fast-changing channel in the symbol timescale to achieve more diversity gain for SM and BF transmissions. Extensive numerical results demonstrate that flexible customization of channel characteristics for a specific transmission scheme can achieve a tradeoff between SE and bit error ratio performance.
\end{abstract}

\section{Introduction}
 {The multiple-input multiple-output (MIMO) technology has been evaluated continuously since the development of wireless communication systems in the 1970s, as reported in research articles on multi-channel digital transmission systems \cite{MIMO-1}--\cite{MIMO-2}.} Depending on the chosen transmission schemes, the MIMO technology can provide multiplexing to increase the transmission rate or diversity to enhance the transmission reliability. Performance tradeoffs between transmission schemes utilizing multiplexing and diversity have been investigated in reference \cite{Tse} from an information theoretic perspective. The suitability and performance of a given MIMO transmission are determined by the channel characteristics.  {Although employing MIMO transmission switching that is adaptive to the channel condition can improve the performance of wireless systems \cite{adaptive-MIMO},} the uncontrollable radio channel restricts the degree of freedom for the design of the transmission scheme. Major performance gains can be expected by breaking the postulate that regards the wireless environment as an uncontrollable element \cite{smart-radio}. The reconfigurable intelligent surface (RIS), which has the ability to reshape electromagnetic waves, is a promising technology to realize the controllable smart radio environment because it is enclosed in the wireless channel between the transmitter (Tx) and the receiver (Rx) \cite{smart-radio-2}.  {Channel customization \cite{channel_cus1}, which designs the RIS to reshape the channel characteristic to cater to the target transmission scheme, can jointly optimize the Tx, wireless channel and the Rx to achieve better system performance.}\par

 {Acquiring channel state information (CSI) is indispensable in realizing the full potential of the RIS in wireless communications. In single-RIS-assisted communication systems, fiber-missing tensor completion tools \cite{estimation-Lin} and atomic norm minimization \cite{estimation-He} were used to reconstruct the CSI by estimating the parameters of the multipath channel. In multi-RIS-assisted systems, the on/off RIS mechanism \cite{estimation-double} and the Bayesian technique \cite{estimation-multi-1} were utilized to address the channel estimation problem. In frequency division duplexing systems, channel feedback is necessary for the base station to obtain the downlink CSI. A novel cascaded codebook was proposed in \cite{bit_par} to adaptively feed back the CSI of the subchannels.  Overhead reduction feedback schemes were proposed for multi-user systems using the sparse characteristic of the cascaded channel \cite{feedback-Shen} \cite{feedback-Shi}. In multi-RIS-assisted systems, the rich-scattering channel was customized to be sparse to reduce the feedback overhead \cite{channel_cus2}. The studies on CSI acquisition have made RIS applications possible.}\par

 {As the RIS provides another degree of freedom for system design, it has been introduced into various wireless applications \cite{App-Han}--\cite{app-dfrc} to improve the system performance. In single-RIS-assisted multi-user communication systems, joint base station (BS) and RIS design was investigated to maximize energy efficiency (EE) \cite{App-Huang} and minimize transmit power \cite{app-Wu}. Multiple RISs were introduced to the cognitive radio system to improve  the spectral efficiency (SE) and EE \cite{App-CR-Liang}. Considering the millimeter wave (mmWave) system assisted by the RIS, a power minimization problem was formulated by jointly optimizing hybrid beamforming at the Tx and the response matrix at the RIS \cite{app-mmwave}. The potential of employing RIS in dual-functional radar-communication systems for improving radar sensing and communication functionalities was investigated in  \cite{app-dfrc}. References \cite{App-Huang}--\cite{app-dfrc} focused on solving the joint design problem for the transceiver and RIS with complex optimization methods and did not explain how the RIS changes the channel. A low-complexity joint Tx--RISs--RX design was proposed in \cite{channel_cus1}, where the RISs' configuration customized the channel characteristic to simplify the optimal singular value decomposition (SVD) transceiver design in hybrid mmWave systems. The channel customization method proposed in \cite{channel_cus1} motivates us to improve the performance of MIMO transmission schemes by directly reshaping the channel rather than using complex optimization methods.} 

 {As obtaining the BS-RIS and RIS-user subchannel matrices in RIS-assisted systems is challenging, two-timescale designs with the statistical and instant CSI were proposed in \cite{MT-Hu}--\cite{MT-Wang}. Exploiting the fact that the BS-RIS subchannel is quasi-static while the RIS-user subchannel is mobile, channel estimation \cite{MT-Hu} and channel feedback \cite{MT-Guo} were performed in the long-term and short-term timescale to simplify the channel acquisition process. In the two-timescale transmission protocol \cite{MT-Zhao}--\cite{MT-Wang}, the passive beamforming at the RIS and the active beamforming at the BS were designed with the statistical CSI in the long-term timescale and the instant CSI in the short-term timescale, respectively.}  {All the aforementioned studies \cite{channel_cus1}--\cite{MT-Wang} designed the RIS in a single timescale, ignoring the fact that the RIS can tailor the channel in different timescales to improve system performance.}\par

 {This study proposes multi-timescale channel customization for RIS-assisted MIMO systems, which avoid the intricate joint Tx-RIS-Rx optimization by dynamically customizing favorable wireless channels for different transmission designs at the Tx and the Rx. Unlike channel customization that aims to simplify the hybrid beamforming design in mmWave systems \cite{channel_cus1} and reduce the feedback overhead in sub-$6$ GHz systems \cite{channel_cus2}, this study attempts to  improve the performance of different transmission schemes in different timescales by proactively reshaping the wireless channel rather than passively adapting to it \cite{adaptive-MIMO}.} The contributions of our work are summarized as follows.\par

\begin{itemize}
\item \textbf{Multi-timescale channel customization framework for transmission design.}  {As the performance of transmission schemes depends on the channel characteristic in different timescales, we introduce the RIS with a shorter reconfigurable time than the symbol time to flexibly customize a suitable channel for the adopted transmission scheme. The proposed framework provides insights into how the customized channel facilitates transmission designs compared to designing transmission schemes without channel customization.}

\item \textbf{Channel customization in the angle coherent time for SE improvement.} Spatial multiplexing (SM) and beamforming (BF) are the effective transmission schemes to increase the SE in the high and low transmit power regimes, respectively. We customize a preferred high-rank channel for the SM transmission and a highly correlated rank-$1$ channel for the BF transmission utilizing the statistical CSI in the
angle coherent time to configure the RIS. Theoretical analysis for the ergodic SE is carried out to compare the relative performance of the channel customization-based SM and BF transmissions.

\item \textbf{Channel customization within the symbol time to increase diversity gain.}  In contrast with the Alamouti space-time code, which forms an equivalent changing and orthogonal channel to obtain the diversity gain, we immediately switch the configuration of the RIS within the symbol time to directly customize a fast-changing channel.The diversity with SM (DS) and the diversity with BF (DB) are proposed with the symbol time level channel customization. We theoretically show that the DS and DB transmission can obtain $M_{\rm R}$-folds diversity by leveraging $M_{\rm R}$-folds SE reduction, where $M_{\rm R}$ is the number of RIS reconfigurations within a symbol time.

\item \textbf{Comprehensive simulations that incorporate insightful discussion about the tradeoff between throughput and reliability.} Numerical results are provided to illustrate the performance of the proposed multi-timescale channel customization for transmission design. The tightness of the closed-form approximation and upper bound derived for the ergodic SE is demonstrated, which presents an intuitive sensing about how the system and environmental parameters, and the channel estimation error affects the ergodic SE. The tradeoff between the throughput and reliability is illustrated by comparing the ergodic SE and the bit error ratio (BER) of the SM, BF, DS, and DB transmission.
\end{itemize}

The rest of this paper is organized as follows. Section \ref{sec:2} describes the system model and the multi-timescale frame structure. Sections \ref{sec:3} and \ref{sec:4} propose the channel customization-based transmission in terms of SE and diversity enhancement, respectively. Section \ref{sec:5} discusses the numerical results. Section \ref{sec:6} provides the conclusions. \par
\emph{Notations}: The lowercase and uppercase of letters denote the vector and matrix, respectively. The transpose and conjugated-transpose operations are represented by superscripts $(\cdot)^T$ and $(\cdot)^H$, respectively; ${\mathbb C}^{a\times b}$ stands for the set of complex $a\times b$ matrix; $|\cdot|$ and $\|\cdot\|$ are used to indicate the absolute value and Euclidean norm, respectively; ${\mathbb E}\{\cdot\}$ calculates the statistical expectation; $[{\bf R}]_{:,i}$ represents the $i$th column of matrix ${\bf R}$. Notation ${\rm blkdiag}\{{\bf X}_1,{\bf X}_2,\,\ldots,\,{\bf X}_N\}$ represents a block diagonal matrix with matrices ${\bf X}_i$, $i=1,\,\ldots,\,N$. Meanwhile ${\rm diag}(a_1,a_2,\,\ldots,\,a_N)$ indicates a diagonal matrix with diagonal elements $a_i$, $i=1, \,\ldots,\, N$.

\section{System Model}\label{sec:2}
\begin{figure}[!t]
\centering
\includegraphics[width=0.5\textwidth]{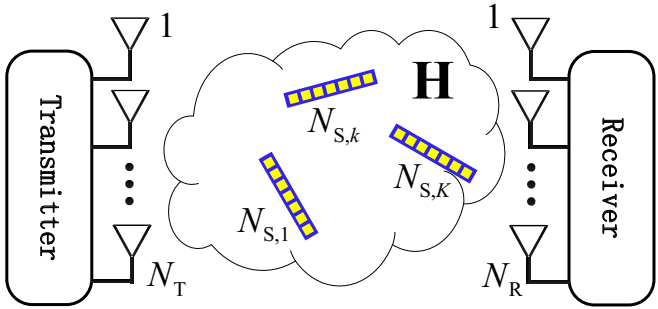}
\caption{MIMO system where the channel consists of multiple RISs.}
\label{Fig.system_model} 
\end{figure}

We consider a time division duplex MIMO system where the direct link between  {the Tx and the Rx} is broken due to the severe blockage\footnote{ {This study considers the scenario without the direct link to exploit the full potential of the proposed multi-timescale channel customization with the common passive RIS. However, note that the active RIS \cite{active-RIS} can reflect and amplify the incident signal, making it possible to extend this work to scenarios where the direct link exists. Additionally, reference \cite{channel_cus1} has shown that by employing sufficient RIS elements, the channel customization method is efficient even in the mmWave system with a strong direct link. }}. $K$ RISs are mounted to establish reliable reflection links to enable data transmission between the Tx and the Rx that locates in the blind coverage without building another Tx. Accordingly, the channel between the Tx and the Rx consists of only components provided by RISs, as shown in Fig. \ref{Fig.system_model}. The channel parameters are assumed to be estimated in the uplink and utilized to configure the RISs to customize the channel for downlink transmission. We assume that the Tx, the Rx, and the RIS are equipped with a uniform linear array (ULA)\footnote{ {If a uniform planar array (UPA) is adopted for the Tx, the Rx, and the RIS, another directional parameter will be introduced for the array response vector. In this case, the array response vector of a UPA can be generated by the array response vector of the horizontal and vertical ULAs. However, since the UPA will not change the main results of this study but complicate the descriptions, we employ the ULA to better focus on the contributions of the multi-timescale channel customization. Note that the single-timescale channel customization with UPA has been investigated in \cite{channel_cus1}.}} with the inter-element spacing being half wavelength. The number of elements in the ULA at the Tx, Rx, and RIS $k$ is denoted by $N_{\rm T}$, $N_{\rm R}$, and $N_{{\rm S},{k}}$, respectively.\par

\subsection{Channel Model}
Given that the direct radio propagation environment from the Tx to the Rx is severely damaged and the reflection links are cascaded by $K$ RISs, the composite channel can be expressed by
\begin{equation}\label{eq:H}
  {\bf{H}} = \sum\nolimits_{k \in {\mathcal K}} {{\rho _k}{{\bf{H}}_{k,{\rm{R}}}}{{\bf{\Gamma }}_k}{{\bf{H}}_{{\rm{T}},k}}},
\end{equation}
where ${\mathcal K}=\{1,\,\ldots,\,K\}$ is the set of RISs; and ${\rho _k} = \frac{\lambda }{{4\pi {r_{{\rm T},k}}}}\frac{\lambda }{{4\pi {r_{k,{\rm{R}}}}}}$ is the cascaded path loss\footnote{The cascaded path loss for the Tx--RIS--Rx channel is the product of the path losses for the Tx--RIS subchannel and the RIS--Rx subchannel, which has been verified by the theoretical and experimental results in \cite{W. Tang_1} and \cite{Emil}.} for the Tx--RIS $k$--Rx link with $\lambda$ being the wavelength and $r_{{\rm T},k}$ and $r_{k,{\rm{R}}}$ being the distances from the Tx and the Rx to the RIS $k$, respectively. In \eqref{eq:H}, matrices ${{\bf{H}}_{k,{\rm{R}}}}\in {{\mathbb C}^{{N_{{\rm{R}}}} \times {N_{{\rm{S}},k}}}}$, ${{\bf{\Gamma }}_k} = {\rm{diag}}( {{e^{j{\omega _{k,1}}}}\,\ldots,\,{e^{j{\omega _{k,{N_{{\rm{S}},k}}}}}}} ) \in {{\mathbb C}^{{N_{{\rm{S}},k}} \times {N_{{\rm{S}},k}}}}$, and ${{\bf{H}}_{{\rm{T}},k}}\in {{\mathbb C}^{{N_{{\rm{S}},k}} \times {N_{{\rm{T}}}}}}$ denote the RIS $k$--Rx channel, the phase shifter response of RIS $k$, and the Tx--RIS $k$ channel, respectively.\par

\begin{figure}[!t]
\centering
\includegraphics[width=0.5\textwidth]{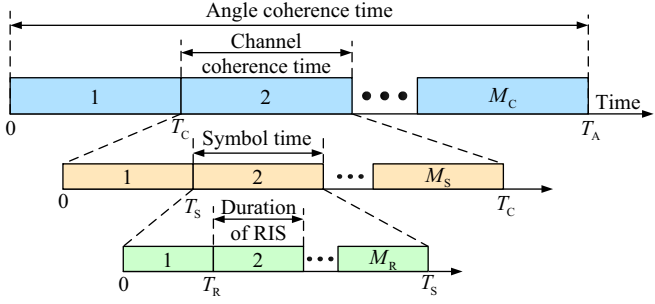}
\caption{Coherent time of parameters in multi-timescale.}
\label{Fig.frame} 
\end{figure}

In a typical system deployment, a line of sight (LoS) link exists in the Tx--RIS $k$ channel to guarantee a stable transmission. Therefore, we adopt the geometry multipath channel model \cite{Saleh} to express the Tx--RIS $k$ channel\footnote{ {The geometry-based multipath channel model is adopted because the phase shifter response of the RIS is sensitive to the incident angle of the EM waves \cite{angle-dependent}. This angle-dependent characteristic of the RIS has no impact on the main results of this study since the phase shifter response is designed only for one path. For simplicity, the environment parameters, such as the Rician factor and the number of paths, are assumed to be identical for each Tx-RIS-Rx link, without loss of generality.}} as
\begin{equation}\label{eq:H_T,k}
\begin{aligned}
  {{\bf{H}}_{{\rm{T}},k}} = \sqrt {{N_{\rm{T}}}{N_{{\rm{S,}}k}}} &\left( \sqrt {\frac{{{\kappa }}}{{{\kappa } + 1}}} {{\bf{a}}_{{\rm{S,}}k}}\left( {\Theta _{k,0}^{\rm{A}}} \right){\bf{a}}_{\rm{T}}^H\left( {\Theta _{k,0}^{\rm{D}}} \right) \right.\\
  &\left. { + \sqrt {\frac{1}{{{\kappa } + 1}}} {\bf{H}}_{{\rm{T}},k}^{{\rm{NLoS}}}} \right),
\end{aligned}
\end{equation}
where ${\kappa }$ is the Rician factor; and  {${{\bf{a}}_{{\rm{S,}}k}}( {\Theta _{k,0}^{\rm{A}}} )$ and ${\bf{a}}_{\rm{T}}( {\Theta _{k,0}^{\rm{D}}} )$} are the array response vectors determined by ${\Theta _{k,0}^{\rm{A}}}=\pi \cos{\theta _{k,0}^{\rm{A}}}$ at the RIS $k$ and ${\Theta _{k,0}^{\rm{D}}}=\pi \cos{\theta _{k,0}^{\rm{D}}}$ at the Tx, respectively, where ${\theta _{k,0}^{\rm{A}}}$ is the angle-of-arrival (AoA), and ${\theta _{k,0}^{\rm{D}}}$ is the angle-of-departure (AoD) of the LoS link. Assuming that the number of non-LoS (NLoS) paths is $L_{{\rm T}}$, the NLoS component of ${{\bf{H}}_{{\rm{T}},k}}$ is given by ${\bf{H}}_{{\rm{T}},k}^{{\rm{NLoS}}} = 1/\sqrt {{L_{{\rm T}}}} \sum\nolimits_{l = 1}^{{L_{{\rm T}}}} {{\beta _{{\rm{T}},k,l}}{{\bf{a}}_{{\rm{S}},k}}( {\Theta _{k,l}^{\rm{A}}} ){\bf{a}}_{\rm{T}}^H( {\Theta _{k,l}^{\rm{D}}} )} $, where $\beta_{{\rm T},k,l}  \sim \mathcal{CN}( 0,1)$ is the fast fading. We denote ${\mathcal L}_{{\rm T}}=\{0,1,\,\ldots,\,L_{{\rm T}}\}$ as the propagation paths for the Tx--RIS $k$ channel and rewrite \eqref{eq:H_T,k} as follows:
\begin{equation}\label{eq:re-H_T,k}
  {{\bf{H}}_{{\rm{T}},k}} = \sum\nolimits_{l \in {\mathcal L}_{{\rm T}}} {{\alpha _{{\rm{T}},k,l}}{{\bf{a}}_{{\rm{S}},k}}\left( {\Theta _{k,l}^{\rm{A}}} \right){\bf{a}}_{\rm{T}}^H\left( {\Theta _{k,l}^{\rm{D}}} \right)} ,
\end{equation}
where ${\alpha _{{\rm{T}},k,0}} =  \sqrt {\frac{{{\kappa }}{N_{\rm{T}}}{N_{{\rm{S,}}k}}}{{{\kappa } + 1}}} $ and ${\alpha _{{\rm{T}},k,l}}=\sqrt {\frac{{{N_{\rm{T}}}{N_{{\rm{S,}}k}}}}{{( {{\kappa } + 1} ){L_{{\rm T},k}}}}} {\beta _{{\rm{T}},k,l}}$ ($l>0$) are the effective path gains for the LoS path and the $l$th NLoS path, respectively. We assume that the statistical CSI (i.e., AoA and AoD) and the instant CSI (i.e., the fast fading) remain constant in angle coherent time $T_{\rm A}$ and channel coherent time $T_{\rm C}$, respectively. Considering that the AoA and AoD change more slowly than the fast fading, the angle coherent time consists of $M_{\rm C}$ channel coherent times, as shown in Fig. \ref{Fig.frame}. In each channel coherent time, $M_{\rm S}$ symbols can be transmitted. The phase shifter of RISs, one of the parameters included in the composite channel can be reconfigured $M_{\rm R}$ times in one symbol time $T_{\rm S}$.\par

 {The Rx is located in a complex environment where the number of scatters and obstacles is large. Unlike the quasi-static Tx--RISs channel, where the Tx and RISs are pre-mounted on high buildings, the RISs--Rx channel may lose the LoS path due to the mobility of the Rx and obstacles. Therefore, the LoS path may not be the dominant component for the RISs--Rx channel. Based on this observation, we omit the Rician factor and merge the LoS path and NLoS paths to express the RIS $k$--Rx channel as follows \cite{feedback-Shi} \cite{sub6-2}:}
\begin{equation}\label{eq:H_R,k}
  {{\bf{H}}_{k,{\rm{R}}}} = \sum\nolimits_{l \in {\mathcal L}_{{\rm{R}}}} {{\alpha _{{k,{\rm{R}}},l}}{{\bf{a}}_{\rm{R}}}\left( {\Phi _{{k},l}^{\rm{A}}} \right){\bf{a}}_{{\rm{S,}}k}^H\left( {\Phi _{{k},l}^{\rm{D}}} \right)},
\end{equation}
where ${\alpha _{{k,{\rm{R}}},l}} = \sqrt {\frac{{{N_{\rm{R}}}{N_{{\rm{S,}}k}}}}{{{L_{{\rm R},k}}}}} {\beta _{{k,{\rm{R}}},l}}$ with $\beta_{{k,{\rm{R}}},l}  \sim \mathcal{CN}( 0,1)$ is the effective path gain for the $l$th path in the RIS $k$--Rx channel, ${\mathcal L}_{{\rm R}}=\{1,\,\ldots,\,L_{{\rm R}}\}$ is the set of paths, and $L_{{\rm R}}$ is the number of paths, ${\bf a}_{\rm R}(\cdot)$ represents the array response vector at the Rx. Meanwhile, ${\Phi _{{k},l}^{\rm{A}}}=\pi \cos{\phi _{{k},l}^{\rm{A}}}$ and ${\Phi _{{k},l}^{\rm{D}}} = \pi \cos{\phi _{{k},l}^{\rm{D}}}$ with ${\phi _{{k},l}^{\rm{A}}}$ and ${\phi _{{k},l}^{\rm{D}}}$ being the AoA and AoD  {of the $l$th path} at the RISs--Rx channel. The array response vectors of the ULA at the Tx,  Rx, and RISs in \eqref{eq:H_T,k}--\eqref{eq:H_R,k} can be uniformly expressed as follows:
\begin{equation}\label{eq:ARV}
  {\bf a}_{X}(Y) = \frac{1}{\sqrt{N_{X}}}{\left[1, e^{jY},\,\ldots,\,e^{j(N_{X}-1)Y}\right]^T},
\end{equation}
where $X\in \{\{{\rm T}\},\{{\rm R}\},\{{\rm S},k\}\}$.\par

The composite channel can be rewritten by substituting \eqref{eq:re-H_T,k} and \eqref{eq:H_R,k} into \eqref{eq:H} and denoting ${\xi _{k,{l},{j}}} \buildrel \Delta \over = {\rho _k}{\alpha _{{k,{\rm{R}}},{l}}}{\alpha _{{\rm{T}},k,{j}}}{\bf{a}}_{{\rm{S,}}k}^H( {\Phi _{{k},{l}}^{\rm{D}}} ){{\bf{\Gamma }}_k}{{\bf{a}}_{{\rm{S}},k}}( {\Theta _{k,{j}}^{\rm{A}}} )$ as the effective path gain of the cascaded path from the $j$th path of the TX--RIS $k$ channel to the $l$th path of the RIS $k$--Rx channel:
\begin{equation}\label{eq:H-3}
  {\bf{H}} = \sum\limits_{k = 1}^K {\sum\limits_{{l} = 1}^{{L_{{\rm R}}}} {\sum\limits_{{j} = 0}^{{L_{{\rm T}}}} {{\xi _{k,{l},{j}}}{{\bf{a}}_{\rm{R}}}\left( {\Phi _{{k},{l}}^{\rm{A}}} \right){\bf{a}}_{\rm{T}}^H\left( {\Theta _{k,{j}}^{\rm{D}}} \right)} } } ={{\bf{R}}}{\bf{\Xi T}}^H,
\end{equation}
where
\begin{equation}\label{eq:M-R}
\begin{aligned}
  {{\bf{R}}} = &\left[ {{\bf{a}}_{\rm{R}}}\left( {\Phi _{{1},1}^{\rm{A}}} \right)\,\ldots,\,{{\bf{a}}_{\rm{R}}}\left( {\Phi _{{1},{L_{{{\rm{R}}}}}}^{\rm{A}}} \right),\,\ldots,\,\right.\\
  &\left.{{\bf{a}}_{\rm{R}}}\left( {\Phi _{{K},1}^{\rm{A}}} \right)\,\ldots,\,{{\bf{a}}_{\rm{R}}}\left( {\Phi _{{K},{L_{{{\rm{R}}}}}}^{\rm{A}}} \right) \right] \in {\mathbb C}^{{N_{\rm R}}\times K{L_{{{\rm{R}}}}}}
\end{aligned}
\end{equation}
and
\begin{equation}\label{eq:M-T}
\begin{aligned}
  {{\bf{T}}} = &\left[ {{{\bf{a}}_{\rm{T}}}\left( {\Theta _{1,0}^{\rm{D}}} \right)\,\ldots,\,{{\bf{a}}_{\rm{R}}}\left( {\Theta _{1,{L_{{\rm{T}}}}}^{\rm{D}}} \right)},\,\ldots,\,\right.\\
  &\left.{{{\bf{a}}_{\rm{T}}}\left( {\Theta _{K,0}^{\rm{D}}} \right)\,\ldots,\,{{\bf{a}}_{\rm{R}}}\left( {\Theta _{K,{L_{{\rm{T}}}}}^{\rm{D}}} \right)} \right]\in {\mathbb C}^{N_{\rm T}\times K(L_{{\rm T}}+1)}
\end{aligned}
\end{equation}
are the array response matrices at the Rx and the Tx; ${\bf{\Xi }} = {\rm{blkdiag}}( {{{\bf{\Xi }}_1}\,\ldots,\,{{\bf{\Xi }}_K}} )\in {\mathbb C}^{KL_{\rm R}\times K(L_{\rm T}+1)}$, where ${{{\bf{\Xi }}_k}} \in {\mathbb C}^{L_{{{\rm{R}}}}\times (L_{{\rm T}}+1)}$ is the effective cascaded path gain matrix for the Tx--RIS $k$--Rx channel with the element in the $l$th row and the $j$th column being ${( {{{\bf{\Xi }}_k}} )_{{l},{j}}} = {\xi _{k,{l},{j}-1}}$.
\subsection{Downlink SE}
When the $m$-stream data ${\bf{s}} = {\left[ {{s_1},{s_2}\,\ldots,\,{s_{m}}} \right]^T} \in {\mathbb C}^{m\times 1}$ satisfying ${\mathbb E}\{{\bf s}{\bf s}^H\}={\bf I}$ is transmitted from the Tx, the received signal at the Rx can be expressed as follows:
\begin{equation}\label{eq:y}
  {\bf{y}} = {{\bf{W}}^H}{\bf{HFs}} + {{\bf{W}}^H}{\bf{n}},
\end{equation}
where ${\bf n}\in {\mathbb C}^{{N_{\rm R}}\times 1}\sim {\mathcal{CN}}(0,\sigma^2 {\bf I})$ is the additive white Gaussian noise with noise power $\sigma^2$; and ${\bf F}\in{\mathbb C}^{{N_{\rm T}}\times{m}}$ and ${\bf W}\in{\mathbb C}^{{N_{\rm R}}\times{m}}$ are the precoder at the Tx and the combiner at the Rx, respectively. The power constraint is $\| {{\bf{Fs}}} \|_F^2 \le E$, where $E$ is the total power at the Tx. Given the received signal, the downlink SE can be expressed as follows:
\begin{equation}\label{eq:R}
  R = {\log _2}\det \left( {{\bf{I}} + \frac{1}{{{\sigma ^2}}}{{\bf{W}}^H}{\bf{HF}}{{\bf{F}}^H}{{\bf{H}}^H}{\bf{W}}} \right),
\end{equation}
which is jointly determined by channel $\bf H$ and transmission schemes $\bf F$ and $\bf W$.\par
In wireless communication systems without the RIS, the design of $\bf F$ and $\bf W$ depends on the uncontrollable $\bf H$. Given that the phase shifter response of the RIS can be regarded as one part of the composite channel itself, introducing RIS to assist wireless communications allows you to customize the channel for specific transmission schemes. Accordingly, the Tx, wireless channel, and the Rx can be jointly designed in the RIS-assisted MIMO systems. Considering that transmission schemes require specific channel characteristics (for example, the SM transmission is supposed to be adopted in the high-rank channel when the system is interference-limited, while the BF transmission is preferred in the rank-deficient channel when the system is noise-limited, and the diversity transmission can obtain more gains in a dynamically changing channel),  {we focus on customizing an expected channel to cater the adopted transmission schemes in different timescales. By customizing the channel structure, the performance of transmission schemes can be improved with simple transceiver designs.}

\section{Channel Customization in Angle Coherent Times for the SM and BF Transmissions}\label{sec:3}
In this section, we first analyze the characteristic of the composite channel and show how channel customization can be realized by configuring the RISs. A full-rank channel and a strongly correlated channel will be customized for the SM and BF transmissions, respectively, by designing the RISs' phase shifter response in the angle coherent time. The ergodic SE of the SM and BF transmissions is analyzed. A transmit power threshold that corresponds to the performance crossing-point is derived based on the analytical results of the ergodic SE to compare the relative SE performance of the SM and BF transmissions.\par

When the RISs are inactive, the phase shifter response ${\bf \Gamma}_k = {\bf I}$ for all $k\in{\mathcal K}$, and the effective path gain of the cascaded path is reduced to
\begin{equation}\label{eq:xi-I}
  {\xi _{k,{l},{j}}}  = {\rho _k}{\alpha _{{k,{\rm{R}}},{l}}}{\alpha _{{\rm{T}},k,{j}}}{\bf{a}}_{{\rm{S,}}k}^H\left( {\Phi _{{k},{l}}^{\rm{D}}} \right){{\bf{a}}_{{\rm{S}},k}}\left( {\Theta _{k,{j}}^{\rm{A}}} \right).
\end{equation}
The RIS is equipped with a mass of elements to ensure that $N_{{\rm S},k}$ is large enough to achieve the asymptotic orthogonality
\begin{equation}\label{eq:aym-orth}
  {\bf{a}}_{{\rm{S,}}k}^H\left( {\Phi _{{k},{l}}^{\rm{D}}} \right){{\bf{a}}_{{\rm{S}},k}}\left( {\Theta _{k,{j}}^{\rm{A}}} \right)\to 0
\end{equation}
when ${\Phi _{{k},{l}}^{\rm{D}}}\ne {\Theta _{k,{j}}^{\rm{A}}}$. The asymptotical orthogonality at the RIS means that if ${\Phi _{{k},{l}}^{\rm{D}}}\ne {\Theta _{k,{j}}^{\rm{A}}}$, inactive RISs result in a poor channel because ${\bf \Xi} \to {\bf 0}$ and ${\bf H}\to {\bf 0}$. To improve the channel quality, we configure the RIS $k$ to align the $j$th AoA and the $l$th AoD on it with the phase shifter response designed by
\begin{equation}\label{eq:optimal-RIS}
  {\omega _{k,n}} = \left( {n - 1} \right)\left( {{\Phi _{{k},{l}}^{\rm{D}}} - {\Theta _{k,{j}}^{\rm{A}}}} \right),
\end{equation}
where $n \in \{ {1\,\ldots,\,{N_{{\rm{S}},k}}} \}$. Accordingly, the cascaded path from the $j$th path of the TX--RIS $k$ channel to the $l$th path of the RIS $k$--Rx channel is activated, and the effective path gain can be maximized as follows:
\begin{equation}\label{eq:xi-max}
  {\xi^{\star} _{k,{l},{j}}}  = {\rho _k}{\alpha _{{k,{\rm{R}}},{l}}}{\alpha _{{\rm{T}},k,{j}}}.
\end{equation}
When the phase shifter response of RIS $k$ is designed as \eqref{eq:optimal-RIS}, we can easily prove that ${\xi _{k,{n},{m}}}\to 0$ still holds for $n\in {\mathcal L}_{\rm R}\backslash\{l\}$  {or} $m\in{\mathcal L}_{\rm T}\backslash \{j\}$. We denote cascaded paths with ${\xi _{k,{n},{m}}}\to 0$ as inactive paths. The composite channel in \eqref{eq:H-3} is then split into two parts as follows:
\begin{equation}\label{eq:H-split}
  {\bf{H}} ={{\bf{R}}}{\bf{\Xi T}}^H={\bf R}_{\rm I}{\bf\Xi}_{\rm I}{\bf T}_{\rm I}^H+{\bf R}_{\rm A}{\bf\Xi}_{\rm A}{\bf T}_{\rm A}^H ,
\end{equation}
where ${\bf{R}}_{\rm I}$ (${\bf{T}}_{\rm I}$) and ${\bf{R}}_{\rm A}$ (${\bf{T}}_{\rm A}$) are array response matrices that consist of the inactive and  {activated} paths at the Rx (Tx), respectively; and ${\bf\Xi}_{\rm I}$ and ${\bf\Xi}_{\rm A}$ contain the corresponding cascaded path gains of the inactive and  {activated} paths, respectively.  {Considering that the inactive paths contribute negligible path gains (i.e., ${\bf\Xi}_{\rm I}\to {\bf 0}$), the composite channel can be approximated as follows \cite{channel_cus2}:
\begin{equation}\label{eq:H-appro}
  {\bf{H}} \approx{\bf R}_{\rm A}{\bf\Xi}_{\rm A}{\bf T}_{\rm A}^H .
\end{equation}}
Subscript $(\cdot)_{\rm A}$ will be replaced by $(\cdot)_{\rm SM}$, $(\cdot)_{\rm BF}$, $(\cdot)_{\rm DS}$, or $(\cdot)_{\rm DB}$ in the rest of this paper, depending on the intended transmission scheme. The approximation in \eqref{eq:H-appro} is determined by the path alignment that is achieved by configuring the RISs' phase shifter in accordance with \eqref{eq:optimal-RIS}. Therefore, the characteristic of the composite channel can be flexibly customized by enhancing the proper cascaded paths with the RISs configuration.\par

In MIMO communications, a high-rank channel is the cornerstone for the SM transmission to increase the SE in the interference-limited regime while a highly correlated channel ensures the BF transmission to gain more signal-to-noise-ratio (SNR) in the noise-limited regime. On the basis of these observations, we design the RISs in different timescales to customize the composite channel that caters to the SM and BF transmissions to improve the SE.

\subsection{Channel Customization for the SM Transmission}\label{Sec:SM}
In this subsection, the statistical CSI is utilized in every angle coherent time to configure the RISs and customize the channel in the form of SVD. With the composite channel being customized into multiple orthogonal sub-channels, the SVD transceiver can be easily designed to obtain multiplexing gains.\par
The path selection and RIS configuration are two key steps for customizing the channel in the form of SVD \cite{channel_cus1} \cite{channel_cus2}. We aim to construct ${\bf R}_{\rm SM}$ and ${\bf T}_{\rm SM}$ that satisfy ${\bf R}_{\rm SM}^H{\bf R}_{\rm SM} ={\bf I}$ and ${\bf T}_{\rm SM}^H{\bf T}_{\rm SM} ={\bf I}$ for \eqref{eq:H-appro} by selecting pair-wise orthogonal paths at the Rx and the Tx. Given that the maximum rank of the composite channel is $N_{\rm R}$, we only select $N_{\rm R}$ paths that are approximately mutually orthogonal. The path selection for ${\bf R}_{\rm SM}\in {\mathbb C}^{N_{\rm R}\times N_{\rm R}}$ at the Rx can be formulated as follows:
\begin{equation}\label{eq:A_R_or}
  \begin{aligned}
{\bf R}_{\rm SM} &= \mathop {\arg \min }\limits_{\bf{R}} \left\| {{{\bf{R}}^H}{\bf{R}} - {\bf{I}}} \right\|_F^2\\
{\rm s. t. }\quad&{\left[ {\bf{R}} \right]_{:,i}} = {{\bf{a}}_{\rm{R}}}\left( {\Phi _{{k_i},{l_i}}^{\rm{A}}} \right),\; i=1,2,\,\ldots,\,N_{\rm R}\\
&k_i \in {{\mathcal K}_ {\rm SM} },\; {{\mathcal K}_ {\rm SM} } \subseteq {\mathcal K},\; \left| {{{\mathcal K}_ {\rm SM} }} \right| = {N_{\rm{R}}},\\
&{l_i} \in {{\mathcal L}_{k_i,{{\rm{R}}}}},\: {{\mathcal L}_{{k_i,{\rm{R}}}}} \subseteq {{\mathcal L}_{{{\rm{R}}}}},\; \left|{{\mathcal L}_{{k_i,{\rm{R}}}}}\right| = 1,
\end{aligned}
\end{equation}
where ${{\mathcal K}_ {\rm SM} }$ is the set of  {activated} RISs for the SM transmission, and ${{\mathcal L}_{{k_i,{\rm{R}}}}}$ is the set of  {activated} paths associated with RIS $k_i$. The path selection in \eqref{eq:A_R_or} is a combinatorial optimization problem and can be straightforwardly addressed by exhaustive search. The path selection at the Tx is similar to that at the RX. We follow references \cite{channel_cus1} and \cite{channel_cus2} to activate the RISs at the DFT directions of the Tx to avoid redundant description for the path selection at the Tx and take full advantage of the fixed position of RISs. Accordingly, the AoD of the LoS path in the Tx--RISs channel can be given by ${\Theta _{k,0}^{\rm{D}}}\in \{\frac{2\pi n}{N_{\rm T}}+\Delta, n=1,2,\,\ldots,\,N_{\rm T}\}$, where $\Delta$ is an arbitrary constant in $[0,2\pi]$. In such deployment, the array response vectors of the LoS path at the Tx are rigorously orthogonal, that is,
\begin{equation}\label{eq:a_T-ortho}
  {\bf{a}}_{\rm{T}}^H\left( {\Theta _{{m},0}^{\rm{D}}} \right){{\bf{a}}_{\rm{T}}}\left( {\Theta _{{n},0}^{\rm{D}}} \right) = 0,m\ne n.
\end{equation}
The array response vectors in $\{{\bf{a}}_{\rm{T}}( {\Theta _{{k},0}^{\rm{D}}}),k\in {\mathcal K}_{\rm SM}\}$ are selected to form ${\bf T}_{\rm SM}\in {\mathbb C}^{N_{\rm T}\times N_{\rm R}}$ on the basis of ${{\mathcal K}_ {\rm SM} }$ obtained from \eqref{eq:A_R_or} and the orthogonality of LoS paths at the Tx. Given that only one cascaded path is picked up to construct ${\bf R}_{\rm SM}$ and ${\bf T}_{\rm SM}$, we configure the  {activated} RISs in ${{\mathcal K}_ {\rm SM} }$ to align the chosen cascaded paths and inactivate the rest of the RISs. The phase shifters of the  {activated} RISs are designed in conformity with \eqref{eq:optimal-RIS} with $k\in {\mathcal K}_{\rm SM}$, $l\in {\mathcal L}_{k,{\rm R}}$, and $j=0$. Now, the composite channel can be approximately customized as follows:
\begin{equation}\label{eq:H-SM}
  {\bf{H}} \approx{\bf R}_{\rm SM}{\bf\Xi}_{\rm SM}{\bf T}_{\rm SM}^H,
\end{equation}
where ${\bf\Xi}_{\rm SM}\in {\mathbb C}^{N_{\rm R}\times N_{\rm R}}$ is a diagonal matrix with elements being the  {activated} effective path gains $\{{\xi^{\star} _{k,{l},{0}}},k\in {\mathcal K}_{\rm SM},l\in{\mathcal L}_{k,{\rm R}}\}$.\par
We have a full row rank channel between the Tx and the Rx with the composite channel customized in the form of SVD as shown in \eqref{eq:H-SM}. To fully exploit the multiplexing gain, $N_{\rm R}$-stream data ${\bf{s}} = [ {{s_1}\,\ldots,\,{s_{{N_{\rm{R}}}}}} ] \in {{\mathbb C}^{{N_{\rm{R}}}\times 1}}$ is transmitted and received with the SVD transceiver
\begin{equation}\label{eq:TXRX-SM}
  {\bf F}_{\rm SM} = \sqrt{\frac{E}{N_{\rm R}}}{\bf T}_{\rm SM}, {\bf W}_{\rm SM} = {\bf R}_{\rm SM}.
\end{equation}
The received signal at the Rx is then given by
\begin{equation}\label{eq:y-SM}
  {\bf{y}} = {\bf{W}}_{{\rm{SM}}}^H{\bf{H}}{\bf F}_{\rm SM}{\bf s} + {\bf{W}}_{{\rm{SM}}}^H{\bf{n}}
\end{equation}
Substituting \eqref{eq:H-SM} and \eqref{eq:TXRX-SM} into \eqref{eq:y-SM} and utilizing ${\bf R}_{\rm SM}^H{\bf R}_{\rm SM} \to{\bf I}$ and ${\bf T}_{\rm SM}^H{\bf T}_{\rm SM} ={\bf I}$ yield
\begin{equation}\label{eq:y-SM-app}
  {\bf{y}} \approx \sqrt {\frac{E}{{{N_{\rm{R}}}}}} {\bf\Xi}_{\rm SM}{\bf{s}} + {\bf{W}}_{{\rm{SM}}}^H{\bf{n}}.
\end{equation}
Given that ${\bf\Xi}_{\rm SM}$ is a diagonal matrix, the $N_{\rm R}$-streams data ${\bf s}$ can be received at the Rx without inter-stream interference, which is the preferred transmission scheme in the interference-limited regime.  {At this point, we can see that because the channel is customized in the form of SVD with a guaranteed rank for the $N_{\rm R}$-streams transmission, the optimal SVD transceiver for the SM transmission can be easily designed without the complex SVD for the channel matrix and the intricate optimization for the joint Tx-RISs-Rx design. Note that we do not claim the optimality of our proposal in terms of  SE maximization because the phase shifter response of the RIS may not be optimal.}

The SE for the SM transmission is then given by
\begin{equation}\label{eq-SE-SM}
  \begin{aligned}
{R_{{\rm{SM}}}} &= {\log _2}\det \left( {{\bf{I}} + \frac{1}{{{\sigma ^2}}}{\bf W}_{\rm SM}^H{\bf{H}}{\bf F}_{\rm SM}{\bf F}_{\rm SM}^H{{\bf{H}}^H}{\bf W}_{\rm SM}} \right)\\
 &\approx \sum\limits_{n = 1}^{{N_{\rm{R}}}} {{{\log }_2}\left( {1 + \frac{{{{\left| {{{\left[ {{{\bf{\Xi}}_{{\rm{SM}}}}} \right]}_{n,n}}} \right|}^2}E}}{{{N_{\rm{R}}}{\sigma ^2}}}} \right)}.
\end{aligned}
\end{equation}
According to \eqref{eq-SE-SM}, the ergodic SE for the SM transmission can be expressed as follows:
\begin{equation}\label{eq:App-SM-1}
\begin{aligned}
  {\bar R_{{\rm{SM}}}} = {\mathbb E}\left\{ {{R_{{\rm{SM}}}}} \right\} &\approx \sum\limits_{n = 1}^{{N_{\rm{R}}}} {\mathbb E}{\left\{ {{{\log }_2}\left( {1 + \frac{{{{\left| {{{\left[ {{{\bf{\Xi}}_{{\rm{SM}}}}} \right]}_{n,n}}} \right|}^2}E}}{{{N_{\rm{R}}}{\sigma ^2}}}} \right)} \right\}} \\
  &= \sum\limits_{n = 1}^{{N_{\rm{R}}}} {\bar R_{{\rm{SM}},n}}.
\end{aligned}
\end{equation}
We ignore the order of the diagonal elements in ${{{\bf{\Xi}}_{{\rm{SM}}}}}$ and let  ${{{[ {{{\bf{\Xi}}_{{\rm{SM}}}}} ]}_{n,n}}}={\xi^{\star} _{n,{l_n},{0}}}$ for $n\in {\mathcal K}_{\rm SM}$ and $l_n\in{\mathcal L}_{n,{\rm R}}$ without loss of generality. The ergodic SE of the $n$th stream data is
\begin{equation}\label{eq:App-SM-2}
\begin{aligned}
  {\bar R_{{\rm{SM}},n}} &=  {\mathbb E}{\left\{ {{{\log }_2}\left( {1 + \frac{{{{\left| {\xi^{\star} _{n,{l_n},{0}}} \right|}^2}E}}{{{N_{\rm{R}}}{\sigma ^2}}}} \right)} \right\}} \\
  &=  {\mathbb E}{\left\{ {{{\log }_2}\left( {1 + \frac{{{{\left| {\rho _n}{\alpha _{{n,{\rm{R}}},{l_n}}}{\alpha _{{\rm{T}},n,{0}}} \right|}^2}E}}{{{N_{\rm{R}}}{\sigma ^2}}}} \right)} \right\}}.
\end{aligned}
\end{equation}
We simplify \eqref{eq:App-SM-2} by expanding ${\alpha _{{n,{\rm{R}}},{l_n}}}$ and ${\alpha _{{\rm{T}},n,{0}}}$:
\begin{equation}\label{eq:App-SM-3}
  {\bar R_{{\rm{SM}},n}} =  {\mathbb E}{\left\{ {{{\log }_2}\left( {1 + \frac{x}{2c_n}} \right)} \right\}},
\end{equation}
where $x=|\sqrt{2}\beta_{n,{\rm R},l_n}|^2$ and $c_n$ is a constant given by
\begin{equation}\label{eq:SM-C-k}
   c_n= {{{\frac{{{\sigma ^2}L_{\rm R}\left( {\kappa  + 1} \right)}}{{E{N_{\rm{T}}}N_{{\rm{S}},n}^2\rho _n^2\kappa }}}}}.
\end{equation}
To explore insights into the SE, two theorems are provided in the following.\par

\begin{theorem}\label{Them:SM-app}
On the basis of the channel customization in \eqref{eq:H-SM} for the SM transmission, the ergodic SE with the SVD transceiver can be approximately given by
\begin{equation}\label{eq-SE-SM-app}
  {{\bar R}_{{\rm{SM}}}} \approx  - \frac{1}{{\ln 2}}\sum\nolimits_{n \in {\mathcal K}_{\rm SM}} {{e^{c_n}}{\rm{Ei}}\left( { - c_n} \right)},
\end{equation}
where ${\rm Ei}(\cdot)$ is the exponential integral defined by reference \cite{Table}:
\begin{equation}\label{eq:Ei}
   {\rm{Ei}}\left( x \right) =  \int\nolimits_{ - \infty }^x {\frac{{{e^t}}}{t}dt} ,{\rm{    }}x < 0.
\end{equation}
\end{theorem}
\begin{IEEEproof}
See Appendix \ref{App:A}.
\end{IEEEproof}

\begin{theorem}\label{Them:SM-upp}
On the basis of the channel customization in \eqref{eq:H-SM} for the SM transmission, the ergodic SE with the SVD transceiver can be upper bounded by
\begin{equation}\label{eq-SE-SM-upp}
\begin{aligned}
  {{\bar R}_{{\rm{SM,upper}}}} &=  \sum\nolimits_{n \in {\mathcal K}_{\rm SM}} {\log_2}\left(1+\frac{1}{c_n}\right) \\
  &= \sum\nolimits_{n \in {\mathcal K}_{\rm SM}} {\log_2}\left(1+\frac{{{E{N_{\rm{T}}}N_{{\rm{S}},n}^2\rho _n^2\kappa }}}{{{{\sigma ^2}L_{\rm R}\left( {\kappa  + 1} \right)}}}\right).
\end{aligned}
\end{equation}
\end{theorem}
\begin{IEEEproof}
Applying Jensen's inequality on the ergodic SE of the $n$th stream data in \eqref{eq:App-SM-3}, we have
\begin{equation}\label{eq:App-SM-u-1}
\begin{aligned}
  {\bar R_{{\rm{SM}},n}} &\le { {{{\log }_2}\left( {1 + \frac{{\mathbb E}\left\{|\beta_{n,{\rm R},l_n}|^2\right\}}{c_n}} \right)} }\\
  &={ {{{\log }_2}\left( {1 + \frac{1}{c_n}} \right)} }={\bar R_{{\rm{SM}},n,{\rm upper}}},
\end{aligned}
\end{equation}
where $c_n$ is given in \eqref{eq:SM-C-k}. ${{\bar R}_{{\rm{SM,upper}}}} =  \sum\nolimits_{n \in {\mathcal K}_{\rm SM}} {\bar R_{{\rm{SM}},n,{\rm upper}}} $ completes the proof.
\end{IEEEproof}

Based on \emph{Theorems \ref{Them:SM-app}} and \emph{ \ref{Them:SM-upp}},  the ergodic SE for the SM transmission depends on $\{c_n, n\in {\mathcal K}_{\rm SM}\}$. Although the approximation of the ergodic SE is complex, the upper bound increases with decreasing $c_n$. Accordingly, we can infer that a larger $\kappa$ or a smaller $L_{\rm R}$ will improve the SE. The underlying reason for this phenomenon is that ${{\bf{H}}_{{\rm{T}},k}}$ and ${{\bf{H}}_{k,{\rm{R}}}}$ are normalized by $\kappa$ and $L_{\rm R}$, respectively, indicating that the cascaded path from the LoS path in the Tx--RIS channel to any one of the paths in the RIS--Rx channel is more significant with a larger $\kappa$ and a smaller $L_{\rm R}$. In this case, the energy leakage reduces, and the SE can be improved when the RIS is configured to enhance this cascaded path. In addition to the environmental parameters, increasing the system parameters (i.e., the transmit power $E$ and the size of antenna arrays $N_{\rm R}$, $N_{\rm T}$, and $N_{{\rm S},k}$) can elevate the SE as well.\par
\subsection{Channel Customization for the BF Transmission}\label{Sec:BF}
In the last subsection, the multiplexing gain is obtained to increase the SE in the interference-limited regime by utilizing the statistical CSI in the angle coherent time to customize a full-row rank channel and design the SVD transceiver. However, the multiplexing is ineffective in the  {noise-limited} regime because the inter-streams interference has fewer impacts on the SE than the noise. Therefore, the BF transmission with single-stream data rather than the SM is preferred in terms of SNR improvement. In this subsection, we customize a strongly correlated rank-$1$ channel to facilitate the BF transmission.\par
To contribute a larger SNR, $K$ RISs are all activated and we customize the channel as follows:
\begin{equation}\label{eq:H-BF}
  {\bf{H}} \approx {\bf R}_{\rm BF}{\bf\Xi}_{\rm BF}{\bf T}_{\rm BF}^H,
\end{equation}
where ${\bf R}_{\rm BF} \in {\mathbb C}^{N_{\rm R}\times K}$, ${\bf\Xi}_{\rm BF}\in {\mathbb C}^{K\times K}$, and ${\bf T}_{\rm BF}\in {\mathbb C}^{N_{\rm T}\times K}$. We still select the LoS paths at the Tx, that is, the columns of ${\bf T}_{\rm BF}$ are  $\{{\bf{a}}_{\rm{T}}( {\Theta _{{k},0}^{\rm{D}}}),k\in {\mathcal K}\}$, for better transmission quality. The path selection for ${\bf R}_{\rm BF}$ and the RISs' design for ${\bf\Xi}_{\rm BF}$ are performed to increase the SNR in the following. When the single-stream data $s$ is transmitted, the precoder that takes full advantage of all  {activated} RIS is designed as follows:
\begin{equation}\label{eq-BF-F}
  {\bf f}_{\rm BF} = \sqrt{\frac{E}{K}}{\bf T}_{\rm BF}{\bm {\eta }}.
\end{equation}
where ${\bm {\eta }}=[\eta_1,\,\ldots,\,\eta_K]^T$ satisfying $|\eta_k|=1$ is the common phase shifter vector and will be designed to increase the SNR in the following. With the combiner ${\bf w}_{\rm BF}$, the received signal at the Rx is given by
\begin{equation}\label{eq:BF-y}
  y  = {\bf w}^H_{\rm BF}{\bf H}{\bf f}_{\rm BF}s + {\bf w}^H_{\rm BF}{\bf n}.
\end{equation}
When the maximum ratio combining (MRC) combiner ${\bf w}_{\rm BF}={\bf H}{\bf f}_{\rm BF}/\| {{\bf{H}}{{\bf{f}}_{{\rm{BF}}}}} \|$ is employed, the received signal can be rewritten as follows:
\begin{equation}\label{eq:BF-y-1}
  y  = \left\| {{\bf{H}}{{\bf{f}}_{{\rm{BF}}}}} \right\|s + {\bf w}^H_{\rm BF}{\bf n}.
\end{equation}
Considering that ${\bf w}^H_{\rm BF}{\bf n}$ does not change the distribution of the noise, the SNR is determined by $\| {{\bf{H}}{{\bf{f}}_{{\rm{BF}}}}} \|$. We denote ${\bf R}_{\rm BF}=[{\bf a}_{{\rm R},1,l_1},\,\ldots,\,{\bf a}_{{\rm R},K,l_K}]$, where ${\bf a}_{{\rm R},k,l_k} = {{\bf{a}}_{\rm{R}}}( {\Phi _{{k},{l_k}}^{\rm{A}}} )$ for $k \in {{\mathcal K} }$ and ${l_k} \in {{\mathcal L}_{{{\rm{R}}}}}$, for clarify of presentation. The received signal can be expressed as follows by substituting \eqref{eq:H-BF} and \eqref{eq-BF-F} into \eqref{eq:BF-y-1}:
\begin{equation}\label{eq:BF-y-2}
\begin{aligned}
  y \approx  &\sqrt{{\frac{E}{K}}\sum\limits_{n = 1}^K {\sum\limits_{m = 1}^K {\eta_n^* \eta_m\left[ {{{\bf{\Xi }}_{{\rm{BF}}}}} \right]_{n,n}^*{{\left[ {{{\bf{\Xi }}_{{\rm{BF}}}}} \right]}_{m,m}}{\bf{a}}_{{\rm{R}},n,{l_n}}^H{{\bf{a}}_{{\rm{R}},m,{l_m}}}} } } s\\
   &+ {\bf{w}}_{{\rm{BF}}}^H{\bf{n}}\\
  \le  &\sqrt{{\frac{E}{K}}\sum\limits_{n = 1}^K {\sum\limits_{m = 1}^K \left| {\left[ {{{\bf{\Xi }}_{{\rm{BF}}}}} \right]_{n,n}^*{{\left[ {{{\bf{\Xi }}_{{\rm{BF}}}}} \right]}_{m,m}}{\bf{a}}_{{\rm{R}},n,{l_n}}^H{{\bf{a}}_{{\rm{R}},m,{l_m}}}} \right| } } s \\
  &+ {\bf{w}}_{{\rm{BF}}}^H{\bf{n}},
\end{aligned}
\end{equation}
where the approximation is due to ${\bf{H}} \approx {\bf R}_{\rm BF}{\bf\Xi}_{\rm BF}{\bf T}_{\rm BF}^H$. The phase of ${\eta_n^* \eta_m[ {{{\bf{\Xi }}_{{\rm{BF}}}}} ]_{n,n}^*{{[ {{{\bf{\Xi }}_{{\rm{BF}}}}} ]}_{m,m}}{\bf{a}}_{{\rm{R}},n,{l_n}}^H{{\bf{a}}_{{\rm{R}},m,{l_m}}}}$ should be aligned for $\forall n,m$ to achieve the equality in \eqref{eq:BF-y-2}. Utilizing the result \cite{bit_par}
\begin{equation}\label{eq:a-H-a}
\begin{aligned}
  {\bf{a}}_{{\rm{R}},n,{l_n}}^H{{\bf{a}}_{{\rm{R}},m,{l_m}}}=& \left| {{\bf{a}}_{{\rm{R}},n,{l_n}}^H{{\bf{a}}_{{\rm{R}},m,{l_m}}}} \right|\\
   &\times {e^{ - j\frac{{{N_{\rm{R}}} - 1}}{2}\Phi _{n,l_n}^{\rm{A}}}} {e^{j\frac{{{N_{\rm{R}}} - 1}}{2}\Phi _{m,l_m}^{\rm{A}}}},
  \end{aligned}
\end{equation}
we have
\begin{equation}\label{eq:APP-BF-4}
  \begin{aligned}
&\eta_n^* \eta_m\left[ {{{\bf{\Xi }}_{{\rm{BF}}}}} \right]_{n,n}^*{\left[ {{{\bf{\Xi }}_{{\rm{BF}}}}} \right]_{m,m}}{\bf{a}}_{{\rm{R}},n,{l_n}}^H{{\bf{a}}_{{\rm{R}},m,{l_m}}}\\
 = &\left| {\left[ {{{\bf{\Xi }}_{{\rm{BF}}}}} \right]_{n,n}^*{{\left[ {{{\bf{\Xi }}_{{\rm{BF}}}}} \right]}_{m,m}}{\bf{a}}_{{\rm{R}},n,{l_n}}^H{{\bf{a}}_{{\rm{R}},m,{l_m}}}} \right|{e^{ - j\angle {\eta_n}}}{e^{ j\angle {\eta_m}}}\\
 &\times{e^{ - j\angle {{\left[ {{{\bf{\Xi }}_{{\rm{BF}}}}} \right]}_{n,n}}}}{e^{j\angle {{\left[ {{{\bf{\Xi }}_{{\rm{BF}}}}} \right]}_{m,m}}}}{e^{ - j\frac{{{N_{\rm{R}}} - 1}}{2}\Phi _{n,{l_n}}^{\rm{A}}}}{e^{j\frac{{{N_{\rm{R}}} - 1}}{2}\Phi _{m,{l_m}}^{\rm{A}}}}.
\end{aligned}
\end{equation}
When ${\angle{\eta_n}}=-\angle {{[ {{{\bf{\Xi }}_{{\rm{BF}}}}} ]}_{n,n}}{- \frac{{{N_{\rm{R}}} - 1}}{2}\Phi _{n,{l_n}}^{\rm{A}}}$, $K^2$ complex variables in \eqref{eq:BF-y-2} can be aligned and the sum is maximized. Given that $[{{{\bf{\Xi }}_{{\rm{BF}}}}}]_{k,k}= {\xi _{k,{l_k},{0}}}  = {\rho _k}{\alpha _{{k,{\rm{R}}},{l_k}}}{\alpha _{{\rm{T}},k,{0}}}{\bf{a}}_{{\rm{S,}}k}^H( {\Phi _{{k},{l_k}}^{\rm{D}}} ){\bf \Gamma}_k{{\bf{a}}_{{\rm{S}},k}}( {\Theta _{k,{0}}^{\rm{A}}} )$, we design ${\bf \Gamma}_k$ in accordance to \eqref{eq:optimal-RIS} with $l=l_k$ and $j=0$ to ensure that $|{\bf{a}}_{{\rm{S,}}k}^H( {\Phi _{{k},{l_k}}^{\rm{D}}} ){\bf \Gamma}_k{{\bf{a}}_{{\rm{S}},k}}( {\Theta _{k,{0}}^{\rm{A}}} )|$ can be maximized. In this case, $\angle {{[ {{{\bf{\Xi }}_{{\rm{BF}}}}} ]}_{k,k}}=\angle {\alpha _{k,{\rm{R}},{l_k}}} + \angle {\alpha _{{\rm{T}},k,{0}}}$,  {$\eta_k$ is then designed as follows}:
\begin{equation}\label{eq:opt-RIS-com}
   e^{-j\left( {\angle {\alpha _{k,{\rm{R}},{l_k}}} + \angle {\alpha _{{\rm{T}},k,{0}}} + \frac{{{N_{\rm{R}}} - 1}}{2}\Phi _{k,l_k}^{\rm{A}}} \right)}
\end{equation}
to reach the equality in \eqref{eq:BF-y-2} by making the received signals superimposed in phase. Considering that a common phase shifter on ${\bf \Gamma}_k$ will not change the result of $|{\bf{a}}_{{\rm{S,}}k}^H( {\Phi _{{k},{l_k}}^{\rm{D}}} ){\bf \Gamma}_k{{\bf{a}}_{{\rm{S}},k}}( {\Theta _{k,{0}}^{\rm{A}}} )|$, an alternative method to reach the equality in \eqref{eq:BF-y-2} is simply setting $\eta_k=1$ and multiplying ${\bf \Gamma}_k$ by \eqref{eq:opt-RIS-com}\footnote{It is noted that the same effect can be achieved by multiplying ${\bf{a}}_{{\rm{R}},k,{l_k}}$ rather than ${\bf \Gamma}_k$ with the the common phase shifter \eqref{eq:opt-RIS-com} when designing ${\bf w}_{\rm BF}$. However, this solution complicates the design of ${\bf w}_{\rm BF}$ because channel path parameters should be estimated at the Rx. In the rich-scattering channel, extracting the path parameters is more complex than directly estimating the effective channel ${\bf H}{\bf f}_{\rm BF}$.}, which is adopted in Section \ref{sec:DB}. Unlike the SM transmission where the transceiver and RISs are designed with a statistical CSI in the angle coherent time, the common phase shifter \eqref{eq:opt-RIS-com} requires the instant CSI in the channel coherent time. Thus, the designing timescale should be taken into account when determining whether the common phase shifter is imposed on the Tx or on the RIS.\par

Given that $|{\bf{a}}_{{\rm{R}},n,{l_n}}^H{{\bf{a}}_{{\rm{R}},m,{l_m}}}| \le 1$,  {\eqref{eq:BF-y-2} can be further rewritten as follows}:
\begin{equation}\label{eq:BF-y-3}
\begin{aligned}
  y=&\sqrt{{\frac{E}{K}}\sum\limits_{n = 1}^K {\sum\limits_{m = 1}^K \left| {\left[ {{{\bf{\Xi }}_{{\rm{BF}}}}} \right]_{n,n}^*{{\left[ {{{\bf{\Xi }}_{{\rm{BF}}}}} \right]}_{m,m}}{\bf{a}}_{{\rm{R}},n,{l_n}}^H{{\bf{a}}_{{\rm{R}},m,{l_m}}}} \right| } } s \\
  &+ {\bf{w}}_{{\rm{BF}}}^H{\bf{n}}\\
\le&  \sqrt{{\frac{E}{K}}\sum\limits_{n = 1}^K {\sum\limits_{m = 1}^K \left| {\left[ {{{\bf{\Xi }}_{{\rm{BF}}}}} \right]_{n,n}^*{{\left[ {{{\bf{\Xi }}_{{\rm{BF}}}}} \right]}_{m,m}}} \right| } } s + {\bf{w}}_{{\rm{BF}}}^H{\bf{n}}.
\end{aligned}
\end{equation}
To approach this equality, the path selection for $\{l_k,k\in {\mathcal K}\}$ should satisfy $|{\bf{a}}_{{\rm{R}},n,{l_n}}^H{{\bf{a}}_{{\rm{R}},m,{l_m}}}| \approx 1$, which can be formulated by
\begin{equation}\label{eq:R-BF}
  \begin{aligned}
{\bf R}_{\rm BF} &= \mathop {\arg \min }\limits_{\bf{R}} \left\| {{{\bf{R}}^H}{\bf{R}} - {\bf{1}}_{K\times K}} \right\|_F^2\\
{\rm s. t. }\quad&{\left[ {\bf{R}} \right]_{:,k}} = {{\bf{a}}_{\rm{R}}}\left( {\Phi _{{k},{l_k}}^{\rm{A}}} \right),\;k \in {{\mathcal K} },\; {l_k} \in {{\mathcal L}_{k,{{\rm{R}}}}},\\
& {{\mathcal L}_{{k,{\rm{R}}}}} \subseteq {{\mathcal L}_{{{\rm{R}}}}},\; \left|{{\mathcal L}_{{k,{\rm{R}}}}}\right| = 1,
\end{aligned}
\end{equation}
where ${\bf{1}}_{K\times K}$ is a full-$1$ matrix.  {The ${\bf R}_{\rm BF}$ and activated paths denoted by $\{{{\mathcal L}_{{k,{\rm{R}}}}},k\in {\mathcal K}\}$ can be obtained by solving \eqref{eq:R-BF} with exhaustive search whose complexity is $L^K_{\rm R}$.} In this case, a strongly correlated rank-$1$ channel is customized because ${\rm{rank}}( {\bf{H}} )={\rm{rank}}( {\bf R}_{\rm BF} )=1$.\par
On the basis of the common phase shifter design \eqref{eq:opt-RIS-com} and path selection \eqref{eq:R-BF}, the equality in \eqref{eq:BF-y-3} can be approached, and the SE for the BF transmission is given as follows:
\begin{equation}\label{eq:SE-BF}
\begin{aligned}
  &R_{\rm BF} = \log_2\left(1+\frac{\left\| {{\bf{H}}{{\bf{f}}_{{\rm{BF}}}}} \right\|^2}{\sigma^2} \right)\\
  \approx &\log_2\left(1+\frac{E {\sum\limits_{n = 1}^K {\sum\limits_{m = 1}^K \left| {\left[ {{{\bf{\Xi }}_{{\rm{BF}}}}} \right]_{n,n}^*{{\left[ {{{\bf{\Xi }}_{{\rm{BF}}}}} \right]}_{m,m}}} \right| } } }{\sigma^2 K} \right).
\end{aligned}
\end{equation}
Considering that the closed-form of the ergodic SE for the BF transmission is difficult to derive due to the absolute operation, we explore insights of the SE by the upper bound given in \emph{Theorem \ref{Them:BF-upp}}.\par
\begin{theorem}\label{Them:BF-upp}
On the basis of the channel customization in \eqref{eq:H-BF} for the BF transmission, the ergodic SE with the MRC receiver can be upper bounded by
\begin{equation}\label{eq:SE-BF-upp}
\begin{aligned}
{{\bar R}_{{\rm{BF,upper}}}} = {\log _2}\left( 1 + \frac{{E{N_{\rm{T}}}{N_{\rm{R}}}\kappa }}{{{\sigma ^2}K{L_{\rm{R}}}\left( {\kappa  + 1} \right)}}\right.\left( \sum\limits_{n = 1}^K {N_{{\rm{S,}}n}^2\rho _n^2} \right. +\\
  \left.\left.\frac{\pi }{4}\sum\limits_{n = 1,n \ne m}^K {\sum\limits_{m = 1}^K {{\rho _n}{\rho _m}{N_{{\rm{S,}}n}}{N_{{\rm{S,}}m}}} }  \right) \right).
\end{aligned}
\end{equation}
\end{theorem}
\begin{IEEEproof}
See Appendix \ref{App:C}.
\end{IEEEproof}\par

We assume $K=N_{\rm R}$ RISs for a fair SE comparison between the BF and the SM transmission without loss of generality. The comparison of \emph{Theorem \ref{Them:BF-upp}} with \emph{Theorem \ref{Them:SM-upp}} demonstrated that $L_{\rm R}$, $\kappa$, $N_{\rm T}$, $E$, and $\sigma^2$ have the same impacts on the upper bound of the ergodic SE for the BF and SM transmission. However, the RISs play different roles in improving the SE. In the SM transmission, the sum operation in ${{\bar R}_{{\rm{SM,upper}}}}$ appears outside the logarithmic function, indicating that the RISs improve the SE by providing more orthogonal sub-channels for data transmission. Meanwhile, the sum operation in ${{\bar R}_{{\rm{BF,upper}}}}$ appears inside of the logarithmic function in the BF transmission. In this case, the RISs are used to increase the SNR at the Rx.\par
\emph{\textbf{Remark 1}}: The different functions of the RISs for the SM and BF transmissions are realized by utilizing the statistical CSI in the angle coherent time to customize different channel characteristics. In the SM transmission, the RISs are configured to create orthogonal paths because the orthogonality, not the correlation is required for the different streams of the data. Although designing the RISs in accordance with \eqref{eq:optimal-RIS} that uses only the statistical CSI can construct a rank-$1$ channel for the BF transmission by redirecting signals to arrive at the Rx from almost the same direction, the signals of different paths may not be coherently superposed. To obtain coherent signals, we refine the precoder or the RISs' configuration with common phase shifters \eqref{eq:opt-RIS-com} and acquire an additional coherent gain. This additional coherent gain is embodied by $\frac{\pi }{4}\sum\nolimits_{n = 1,n \ne m}^K {\sum\nolimits_{m = 1}^K {{\rho _n}{\rho _m}{N_{{\rm{S,}}n}}{N_{{\rm{S,}}m}}} }$ in \emph{Theorem \ref{Them:BF-upp}} and will be replaced by zero when the common phase shifter \eqref{eq:opt-RIS-com} is not applied.\par
\emph{\textbf{Remark 2}}: The SM transmission that transmits multi-stream data via single-orthogonal paths and the BF transmission that transmits single-stream data via multi-correlated paths are two special cases for the proposed channel customization. The discussed SM and BF transmissions focus on solely obtaining the multiplexing and SNR gains, respectively. In typical cases, the multiplexing and SNR gains can be simultaneously obtained when the channel is properly customized for transmission schemes. For example, additional $K-N_{\rm R}$ RISs can be selected for the SM transmission to provide SNR gain by redirecting the extra paths strongly correlated to the orthogonal paths. In the BF transmission, multiplexing gain can be obtained by reshaping some paths out from the strongly correlated ones to be orthogonal. The coupling between the multiplexing and SNR gains depends on the path selection strategies and is left for our future work due to the limited space.

\subsection{Performance Crossing-point for the SM and BF Transmissions}\label{sec:CP}
In the preceding subsections, we customize the channel for the SM transmission in the interference-limited regime and for the BF transmission in the noise-limited regime to improve the SE.   {In this subsection, we investigate the preferred transmit power regimes for SM and BF transmissions by deriving the performance crossing-point of the ergodic SE upper bounds. We use the upper bounds, rather than exact results or approximations, as they provide explicit insights into how system parameters determine the ergodic SE. While the resulting performance crossing-point may be biased, it will not change the preferred transmit power regimes for SM and BF transmissions.}\par
To theoretically compare the performance of the SM and BF transmissions, the upper bounds must be manipulated into slightly different forms. Following \cite[Eqs.$(44)$--$(47)$]{adaptive-MIMO}, the upper bound of the SM transmission can be rewritten as follows:
\begin{equation}\label{eq-SE-SM-upp-1}
\begin{aligned}
  {{\bar R}_{{\rm{SM}},{\rm{upper}}}} &= {\log _2}\left( {\prod\limits_{n = 1}^{{N_{\rm{R}}}} {\left( {1 + \frac{{E{N_{\rm{T}}}N_{{\rm{S}},n}^2\rho _n^2\kappa }}{{{\sigma ^2}{L_{\rm{R}}}\left( {\kappa  + 1} \right)}}} \right)} } \right)\\
   &= {\log _2}\left( {1 + \sum\limits_{n = 1}^{{N_{\rm{R}}}} {{{\left( {\frac{{E{N_{\rm{T}}}\kappa }}{{{\sigma ^2}{L_{\rm{R}}}\left( {\kappa  + 1} \right)}}} \right)}^n}{{\rm{tr}}_n}\left( {{{\bf{D}}_{{N_{\rm{R}}}}}} \right)} } \right),
\end{aligned}
\end{equation}
where
\begin{equation}\label{eq:D_N_R}
  {{\bf{D}}_{{N_{\rm{R}}}}} = {\rm{diag}}\left( {N_{{\rm{S}},1}^2\rho _1^2\,\ldots,\,N_{{\rm{S}},{N_{\rm{R}}}}^2\rho _{{N_{\rm{R}}}}^2} \right),
\end{equation}
and ${{\rm{tr}}_n}(\cdot)$ denotes the $n$th symmetric function,
which is defined as follows:
\begin{equation}\label{eq:tr_k}
  {\rm{t}}{{\rm{r}}_k}\left( {\bf{A}} \right) = \sum\limits_{\left\{ {\vec{a}} \right\}}{\prod\limits_{i = 1}^{k}{\lambda_{x,a_i}}}=\sum\limits_{\left\{ {\vec{a}} \right\}} {\left| {{{\left[ {\bf{A}} \right]}_{{\vec{a}},{\vec{a}}}}} \right|}
\end{equation}
for arbitrary Hermitian positive-definite ${\bf A}\in {\mathbb C}^{N\times N}$. In \eqref{eq:tr_k}, the sum is over all ordered ${\vec{a}}=\{a_1,\,\ldots,\,a_k\}\subseteq \{1,\,\ldots,\,N\}$, $\lambda_{x,i}$ denotes the $i$th eigenvalue of ${\bf A}$, and ${{{\left[ {\bf{A}} \right]}_{{\vec{a}},{\vec{a}}}}}$ is the $k\times k$ principle sub-matrix of ${\bf A}$ constructed by taking only the rows and columns indexed by ${\vec{a}}$. With the definition in \eqref{eq:tr_k}, we have
\begin{equation}\label{eq:tr_1}
  {\rm{t}}{{\rm{r}}_1}\left( {{{\bf{D}}_{K}}} \right) = \sum\limits_{n = 1}^{K} {N_{{\rm{S,}}n}^2\rho _n^2} ,
\end{equation}
and
\begin{equation}\label{eq:tr_2}
  {\rm{t}}{{\rm{r}}_2}\left( {{\bf{D}}_{K}^{1/2}} \right) = \sum\limits_{n = 1}^{K} {\sum\limits_{m >n}^{K} {{\rho _n}{\rho _m}{N_{{\rm{S,}}n}}{N_{{\rm{S,}}m}}} }.
\end{equation}
Substituting \eqref{eq:tr_1} and \eqref{eq:tr_2} into \eqref{eq:SE-BF-upp} yields
\begin{equation}\label{eq:SE-BF-upp-1}
\begin{aligned}
{{\bar R}_{{\rm{BF,upper}}}} = &{\log _2}\left( 1 + \frac{{E{N_{\rm{T}}}{N_{\rm{R}}}\kappa }}{{{\sigma ^2}K{L_{\rm{R}}}\left( {\kappa  + 1} \right)}}\right.\times\\
&\left.\left( {{\rm{t}}{{\rm{r}}_1}\left( {{{\bf{D}}_{K}}} \right)  + \frac{\pi }{2}{\rm{t}}{{\rm{r}}_2}\left( {{\bf{D}}_{K}^{1/2}} \right) } \right) \right).
\end{aligned}
\end{equation}
When the SM and BF transmissions have the crossing-point on the upper bound of the ergodic SE (i.e., ${{\bar R}_{{\rm{SM,upper}}}} = {{\bar R}_{{\rm{BF,upper}}}}$), we can obtain
\begin{equation}\label{eq:upper-equal}
\begin{aligned}
  1 &+ {{{\left( {\frac{{E{N_{\rm{T}}}\kappa }}{{{\sigma ^2}{L_{\rm{R}}}\left( {\kappa  + 1} \right)}}} \right)}}{{\rm{tr}}_1}\left( {{{\bf{D}}_{{N_{\rm{R}}}}}} \right)} \\
  &+ \sum\limits_{n = 2}^{{N_{\rm{R}}}} {{{\left( {\frac{{E{N_{\rm{T}}}\kappa }}{{{\sigma ^2}{L_{\rm{R}}}\left( {\kappa  + 1} \right)}}} \right)}^n}{{\rm{tr}}_n}\left( {{{\bf{D}}_{{N_{\rm{R}}}}}} \right)} \\
  =1 &+ \frac{{E{N_{\rm{T}}}{N_{\rm{R}}}\kappa }}{{{\sigma ^2}K{L_{\rm{R}}}\left( {\kappa  + 1} \right)}}\left( {{\rm{t}}{{\rm{r}}_1}\left( {{{\bf{D}}_{K}}} \right)  + \frac{\pi }{2}{\rm{t}}{{\rm{r}}_2}\left( {{\bf{D}}_{K}^{1/2}} \right) } \right).
\end{aligned}
\end{equation}
Therefore, the transmit power threshold $E_{\rm th}$ corresponding to the SE crossing-point is given by the positive solution to the polynomial equation
\begin{equation}\label{eq:poly-equ}
\begin{aligned}
&\sum\limits_{n = 2}^{{N_{\rm{R}}}} {{{\left( {\frac{{E{N_{\rm{T}}}\kappa }}{{{\sigma ^2}{L_{\rm{R}}}\left( {\kappa  + 1} \right)}}} \right)}^{n - 1}}{\rm{t}}{{\rm{r}}_n}\left( {{{\bf{D}}_{{N_{\rm{R}}}}}} \right)} \\
  =&\frac{{{N_{\rm{R}}}}}{K}\left( {{\rm{t}}{{\rm{r}}_1}\left( {{{\bf{D}}_K}} \right) + \frac{\pi }{2}{\rm{t}}{{\rm{r}}_2}\left( {{\bf{D}}_K^{1/2}} \right)} \right) - {\rm{t}}{{\rm{r}}_1}\left( {{{\bf{D}}_{{N_{\rm{R}}}}}} \right).
\end{aligned}
\end{equation}
We now provide closed-form expressions for $E_{\rm th}$ for two special cases to gain further insight.
\begin{enumerate}
  \item \emph{$N_{\rm R}=2$: }In this case, the solution to \eqref{eq:poly-equ} is given by
  \begin{equation}\label{eq:N_R-2-1}
  \begin{aligned}
    {E_{{\rm{th}}}} & = \frac{{{\sigma ^2}{L_{\rm{R}}}\left( {\kappa  + 1} \right)}}{{{N_{\rm{T}}}\kappa }}\\
    &\times \frac{{\frac{2}{K}\left( {{\rm{t}}{{\rm{r}}_1}\left( {{{\bf{D}}_K}} \right) + \frac{\pi }{2}{\rm{t}}{{\rm{r}}_2}\left( {{\bf{D}}_K^{1/2}} \right)} \right) - {\rm{t}}{{\rm{r}}_1}\left( {{{\bf{D}}_2}} \right)}}{{{\rm{t}}{{\rm{r}}_2}\left( {{{\bf{D}}_2}} \right)}}.
    \end{aligned}
  \end{equation}
  When each RIS has a comparative number of elements to combat the path loss to ensure that ${N_{{\rm{S,}}n}\rho _n}$ is a constant denoted by $C$, \eqref{eq:N_R-2-1} can be reduced to
  \begin{equation}\label{eq:N_R-2-2}
    {E_{{\rm{th}}}}  = \frac{{{\sigma ^2}{L_{\rm{R}}}\left( {\kappa  + 1} \right)\pi \left( {K - 1} \right)}}{{2{N_{\rm{T}}}\kappa {C^2}}}.
  \end{equation}
$C$ represents the scale of the RIS because it is proportional to $N_{{\rm S},n}$ when the path loss is fixed. Equation \eqref{eq:N_R-2-2} shows that the parameters that determine the transmit power threshold have an opposite influence on $E_{\rm th}$ than on  ${{\bar R}_{{\rm{SM,upper}}}}$ and ${{\bar R}_{{\rm{BF,upper}}}}$. The transmit power threshold  inversely varies to the number of transmit antennas and RIS's elements, indicating that the relative SE improvement benefiting from the array gains at the Tx and RISs is greater for the SM transmission than for the BF transmission. We see that $E_{\rm th}$ increases with $K$ because more SNR gains can be obtained for the BF transmission when more RISs are involved to customize the strongly correlated rank-$1$ channel.

  \item \emph{$N_{\rm R}=3$: }In this case, \eqref{eq:poly-equ} reduces to a simple quadratic equation. The positive solution to the quadratic equation is given by \eqref{eq:N_R-3-1}.
      \begin{figure*}
      \begin{equation}\label{eq:N_R-3-1}
       {E_{{\rm{th}}}}= \frac{{{\sigma ^2}{L_{\rm{R}}}\left( {\kappa  + 1} \right)}}{{{N_{\rm{T}}}\kappa }}\frac{{\sqrt {{\rm{t}}{{\rm{r}}_2}{{\left( {{{\bf{D}}_3}} \right)}^2} - 4{\rm{t}}{{\rm{r}}_3}\left( {{{\bf{D}}_3}} \right)\left( {{\rm{t}}{{\rm{r}}_1}\left( {{{\bf{D}}_3}} \right) - \frac{3}{K}\left( {{\rm{t}}{{\rm{r}}_1}\left( {{{\bf{D}}_K}} \right) + \frac{\pi }{2}{\rm{t}}{{\rm{r}}_2}\left( {{\bf{D}}_K^{1/2}} \right)} \right)} \right)}  - {\rm{t}}{{\rm{r}}_2}\left( {{{\bf{D}}_3}} \right)}}{{2{\rm{t}}{{\rm{r}}_3}\left( {{{\bf{D}}_3}} \right)}}
      \end{equation}
      \hrulefill
      \end{figure*}
      When ${N_{{\rm{S,}}n}\rho _n}=C$, this solution is reduced to
      \begin{equation}\label{eq:N_R-3-s}
       {E_{{\rm{th}}}} = \frac{{{\sigma ^2}{L_{\rm{R}}}\left( {\kappa  + 1} \right)}}{{{N_{\rm{T}}}\kappa }}\frac{{\sqrt {9 + 3\pi \left( {K - 1} \right)}  - 3}}{{2{C^2}}}
      \end{equation}
      Variables $\sigma$, $L_{\rm R}$, ${N_{\rm T}}$, $K$, and $\kappa$ have the same impact on $E_{\rm th}$ for $N_{\rm R}=3$ as for $N_{\rm R}=2$. To explore how $E_{\rm th}$ varies with $N_{\rm R}$, we compare \eqref{eq:N_R-3-s} with \eqref{eq:N_R-2-2} and have
      \begin{equation}\label{eq:N_R-3-2}
      \begin{aligned}
      &{E_{{\rm{th}}}}\Big{|}_{N_{\rm R}=3}-{E_{{\rm{th}}}}\Big{|}_{N_{\rm R}=2}\\
       =&\frac{{{\sigma ^2}{L_{\rm{R}}}\left( {\kappa  + 1} \right)}}{{2{N_{\rm{T}}}\kappa {C^2}}}\left( {\sqrt {9 + 3\pi \left( {K - 1} \right)} -   {K\pi}+ \pi - 3 } \right)\\
       \mathop  \le \limits^{\left( a \right)}& 0,
       \end{aligned}
      \end{equation}
      where $(a)$ is because ${\sqrt {9 + 3\pi \left( {K - 1} \right)} -   {K\pi} } $ is a monotone decreasing function of $K$ with the maximum value being zero. According to this result, the increasing array gain at the Rx also brings a greater relative SE improvement for the SM transmission than for the BF transmission.
\end{enumerate}

\section{Channel Customization within Symbol Times for Diversity}\label{sec:4}
Using MIMO techniques to increase the transmission rate or enhance the transmission reliability depends largely on the channel characteristic. In the preceding section, the composite channel is customized for the SM and BF transmissions to improve the SE. In this section, we focus on enhancing the transmission reliability by customizing the composite channel to obtain more diversity gains. The basic principle is that diversity gain can be achieved in a fast-changing channel.\par
 {The Alamouti code \cite{alamouti} is a technique that can provide diversity gain and eliminate signal interference in a changeless channel and is still widely used in 5G to create an equivalently varying and orthogonal channel \cite{alamouti_GFDM} \cite{alamouti_5G}. However, for wireless systems without RISs, Alamouti utilizes the complex orthogonal space time code to form the preferred channel for diversity, but it is difficult to find a complex space-time code that achieves the maximum diversity gain and the maximum coding rate at the same time when the transmitter has more than two antennas \cite{Y. S. Cho}. With the introduction of RISs with a shorter configuration time than the channel coherent time,} the composite channel itself can be customized to be time-varying with different characteristics, which avoids the limitation of the space-time code and opens a new research direction on diversity.  {For example, the time diversity obtained by retransmission \textbf{between} the channel coherent times is unpopular in some scenarios, such as ultra-reliable low-latency communications that require low latency. However, by retransmitting consecutive symbols with the RIS switching between symbol times, time diversity can be obtained, improving reliability \textbf{within} the channel coherent time because the composite channel is customized to be different for each symbol. Furthermore, even \textbf{without} retransmission, diversity gain can be achieved by multi-reception when the RIS is quickly reconfigured within the symbol time, making it appealing for low-latency applications. The diversity gained by reshaping the wireless channel with RISs is worth investigating.} Thus, we pursue the diversity gain within the channel coherent time for the SM and BF transmissions in the following subsections.

\subsection{Diversity with the SM}\label{sec:DS}
Inspired by the Alamouti code and the channel customization, the diversity gain and interference-elimination can be achieved by directly constructing orthogonal channels within symbol time. In Fig. \ref{Fig.frame}, the phase shifter of the RIS can be reconfigured $M_{\rm R}$ times within a symbol time. During the $m$th configuration, we assume that the composite channel can be customized as follows by utilizing the path selection and the RISs' phase shifter design:
\begin{equation}\label{eq:H-DS-m}
  {\bf{H}}_m \approx{\bf R}_{{\rm DS},m}{\bf\Xi}_{{\rm DS},m}{\bf T}_{{\rm DS}}^H,
\end{equation}
where ${\bf R}_{{\rm DS},m}\in {\mathbb C}^{N_{\rm R}\times N_{\rm R}}$ satisfying ${\bf R}_{{\rm DS},m}^H{\bf R}_{{\rm DS},m}\to {\bf I}$ is selected in accordance with \eqref{eq:A_R_or}; ${\bf\Xi}_{{\rm DS},m}\in {\mathbb C}^{N_{\rm R}\times N_{\rm R}}$ is a diagonal matrix formed by designing $ {{{\bf{\Gamma }}_{k,m}}} $ according to \eqref{eq:optimal-RIS} with $k\in {\mathcal K}_{{\rm DS}}$, $l\in {\mathcal L}_{k,{\rm R},m}$, and $j=0$, where ${\mathcal K}_{{\rm DS}}$ and $\{{\mathcal L}_{k,{\rm R},m},k\in {\mathcal K}_{{\rm DS}} \}$ are the byproducts of the path selection; and ${\bf T}_{{\rm DS}}\in {\mathbb C}^{N_{\rm T}\times N_{\rm R}}$ satisfying ${\bf T}_{{\rm DS}}^H{\bf T}_{{\rm DS}}= {\bf I}$ is constructed by using the array response vectors of the LoS path in the Tx--RIS $k$ ($k\in {\mathcal K}_{{\rm DS}}$) channel\footnote{It is noted that we only select ${\mathcal K}_{{\rm DS}}$ once because the corresponding ${\bf T}_{{\rm DS}}$ that is used as the precoder at the Tx should be constant for a transmitted symbol.}. In $M_{\rm R}$ RISs' reconfigurations, we select different paths from the RISs--Rx channel (i.e., ${ {{\cap}} _{m\in \{1,\,\ldots,\,M_{\rm R}\}}}{{\mathcal L}_{k,{\rm{R}},m}}=\emptyset$) to ensure diversity. Although the constraint on path selection increases the diversity gain, it inevitably deteriorates the orthogonality of ${\bf R}_{{\rm DS},m}$. This phenomenon embodies the tradeoff between the reliability and the throughput of wireless communication systems.\par

The precoder at the Tx and the combiner at the Rx for the $m$th RIS configuration are designed as follows on the basis of the channel customization result in \eqref{eq:H-DS-m}:
\begin{equation}\label{eq:TXRX-DS}
  {\bf F}_{\rm DS} = \sqrt{\frac{E}{N_{\rm R}}}{\bf T}_{\rm DS}, {\bf W}_{{\rm DS},m} = {\bf R}_{{\rm DS},m}.
\end{equation}
The signals that arrive at the Rx are given by
\begin{equation}\label{eq:r-DS-m}
  {\bf{r}}_m = {\bf{H}}_m{\bf F}_{\rm DS}{\bf s} + {\bf{n}}_m.
\end{equation}
Stacking all signals for $M_{\rm R}$ RIS configuration yields:
\begin{equation}\label{eq:r-DS}
  {\bf{r}} = \left[{\bf{r}}_1,\,\ldots,\,{\bf{r}}_{M_{\rm R}}\right]^T = {\bf{H}}_{\rm DS}{\bf F}_{\rm DS}{\bf s} + {\bf{n}},
\end{equation}
where ${\bf{H}}_{\rm DS}=[{\bf{H}}_1,\,\ldots,\,{\bf{H}}_{M_{\rm R}}]^T$ and ${\bf{n}}=[{\bf{n}}_1,\,\ldots,\,{\bf{n}}_{M_{\rm R}}]^T$. In the ideal case where ${\bf R}_{{\rm DS},m}^H{\bf R}_{{\rm DS},m}\to {\bf I}$, we multiply ${\bf r}$ by ${\bf{W}}_{{\rm{DS}}}=[{\bf{W}}_{{\rm{DS}},1}^H,\,\ldots,\,{\bf{W}}_{{\rm{DS}},M_{\rm R}}^H]^H$ and have
\begin{equation}\label{eq:y_DS}
  \begin{aligned}
{\bf{ y}} &= {\bf{W}}_{{\rm{DS}}}^H{\bf{r}} = {\bf{W}}_{{\rm{DS}}}^H{{\bf{H}}_{{\rm{DS}}}}{{\bf{F}}_{{\rm{DS}}}}{\bf{s}} + {\bf{W}}_{{\rm{DS}}}^H{\bf{n}}\\
 &\approx \sqrt {\frac{E}{{{N_{\rm{R}}}}}} \sum\limits_{m = 1}^{{M_{\rm{R}}}} {{{\bf{\Xi}}_{{\rm{DS}},m}}} {\bf{s}} + {\bf{W}}_{{\rm{DS}}}^H{\bf{n}}.
\end{aligned}
\end{equation}
Given the path selection constraint ${ {{\cap}} _{m\in \{1,\,\ldots,\,M_{\rm R}\}}}{{\mathcal L}_{k,{\rm{R}},m}}=\emptyset$, $\sum\nolimits_{m = 1}^{{M_{\rm{R}}}}[{{{\bf{\Xi}}_{{\rm{DS}},m}}}]_{k,k} $ warrants the diversity order of $M_{\rm R}$ for ${ s}_k$ because $[{{{\bf{\Xi}}_{{\rm{DS}},m}}}]_{k,k}\ne [{{{\bf{\Xi}}_{{\rm{DS}},n}}}]_{k,k}$ for $m\ne n$. To improve the reception quality of $s_k$, we multiply $\{{\bf \Gamma}_{k,m}\}_{m=1}^{M_{\rm R}}$ or $\{[{\bf{W}}_{{\rm{DS}},k}]_{:,m}\}_{m=1}^{M_{\rm R}} $ with common phase shifters similar to \eqref{eq:opt-RIS-com} to ensure that  $\{[{{{\bf{\Xi}}_{{\rm{DS}},m}}}]_{k,k}\}_{m=1}^{M_{\rm R}}$ can be coherently added. Given that ${\mathbb E}\{{\bf{W}}_{{\rm{DS}}}^H{\bf{n}}{\bf{n}}^H{\bf{W}}_{{\rm{DS}}}\}={M_{\rm R}}\sigma^2{\bf I}$, the noise power of each stream data is enlarged, and the outage probability of the $k$th stream data is given by
\begin{equation}\label{eq-OP-DS}
{{ P}_{{\rm{DS}},k}} = \left\{\frac{{{\left(\sum\limits_{m = 1}^{{M_{\rm{R}}}}\left|[{{{\bf{\Xi}}_{{\rm{DS}},m}}}]_{k,k}\right|\right)^2}E}}{{{N_{\rm{R}}}{M_{\rm{R}}}{\sigma ^2}}}<\gamma_{\rm th}\right\},
\end{equation}
where $\gamma_{\rm th}$ is the SNR threshold and the absolute operation is derived from the refined $\{{\bf \Gamma}_{k,m}\}_{m=1}^{M_{\rm R}}$ or $\{[{\bf{W}}_{{\rm{DS}},k}]_{:,m}\}_{m=1}^{M_{\rm R}} $. When $M_{\rm R}=1$, the DS transmission is reduced to the SM transmission. Accordingly, the outage probability of the SM transmission, ${{ P}_{{\rm{SM}}}}$, is a special case of ${{ P}_{{\rm{DS}}}}$ with $M_{\rm R}=1$. Although the closed-form expression for ${{ P}_{{\rm{DS}}}}$ is difficult to obtain, we can easily observe that the received signal power and the noise power  quadratically and linearly increase to $M_{\rm R}$, respectively. These scaling laws indicate that the outage probability of the DS transmission is lower than the SM transmission and the reliability is enhanced given a fixed $\gamma_{\rm th}$. The reliability improvement is at the cost of the SE reduction. The SE for the DS transmission is given by:
\begin{equation}\label{eq-SE-DS}
{{ R}_{{\rm{DS}}}} \approx \frac{1}{M_{\rm R}}\sum\limits_{n = 1}^{{N_{\rm{R}}}} {{{\log }_2}\left( 1 + \frac{{{\left(\sum\limits_{m = 1}^{{M_{\rm{R}}}}\left|[{{{\bf{\Xi}}_{{\rm{DS}},m}}}]_{n,n}\right|\right)^2}E}}{{{N_{\rm{R}}}{M_{\rm{R}}}{\sigma ^2}}} \right)}.
\end{equation}
In the comparison of  ${{ R}_{{\rm{DS}}}}$ with ${{ R}_{{\rm{SM}}}}$, the DS transmission provides a diversity gain of $M_{\rm R}$ on the SNR while reduces the SE $M_{\rm R}$-folds, indicating the tradeoff between reliability and throughput. Furthermore, the coherent signal superposition brings additional SNR gain for the DS transmission. This additional SNR gain for the DS transmission is similar to the BF when compared with the SM. Thus, we anticipate an SE crossing-point for the DS and SM transmissions, which is similar to that for the BF and SM transmissions.\par

In comparison with enhancing the reliability with a wider transmission bandwidth in wireless communication systems without RISs, the DS transmission has advantages in interference elimination because orthogonal subchannels can be customized.
\subsection{Diversity with the BF}\label{sec:DB}
In Section \ref{sec:DS}, we customize the channel within symbol times for the DS transmission to achieve the diversity gain on the basis of the SM transmission. However, subject to the multiplexing gain, the diversity order is limited to $M_{\rm R}$. From the perspective of reliability, a diversity order of $K^2$ is achieved by the BF transmission at the cost of multiplexing gain, as shown in \eqref{eq:BF-y-2}. The diversity gain in the BF transmission is obtained from the spatial diversity and can be further improved from the time diversity in the DB transmission.\par

We reconfigure the RIS's phase shifters $M_{\rm R}$ times within a symbol time following the process in the DS transmission. During the $m$th reconfiguration, the composite channel is customized as follows:
\begin{equation}\label{eq:H-DB-m}
  {\bf{H}}_m \approx{\bf R}_{{\rm DB},m}{\bf\Xi}_{{\rm DB},m}{\bf T}_{{\rm DB}}^H,
\end{equation}
where ${\bf R}_{{\rm DB},m}\in {\mathbb C}^{N_{\rm R}\times K}$ satisfying ${\bf R}_{{\rm DB},m}^H{\bf R}_{{\rm DB},m}\to {\bf 1}_{K\times K}$ is selected in accordance with \eqref{eq:R-BF}; ${\bf\Xi}_{{\rm DB},m}\in {\mathbb C}^{K\times K}$ is a diagonal matrix formed by designing $ {{{\bf{\Gamma }}_{k,m}}} $ according to \eqref{eq:optimal-RIS} with $k\in {\mathcal K}_{{\rm DB}}$, $l\in {\mathcal L}_{k,{\rm R},m}$, and $j=0$, where ${\mathcal K}_{{\rm DB}}$ and $\{{\mathcal L}_{k,{\rm R},m},k\in {\mathcal K}_{{\rm DB}} \}$ are the byproducts of the path selection; and ${\bf T}_{{\rm DB}}\in {\mathbb C}^{N_{\rm T}\times K}$ satisfying ${\bf T}_{{\rm DB}}^H{\bf T}_{{\rm DB}}= {\bf I}$ is constructed by array response vectors of the LoS path in the Tx--RIS $k$ ($k\in {\mathcal K}_{{\rm DB}}$) channel. In the DB transmission, we also select different paths for $M_{\rm R}$ RISs' reconfiguration to achieve diversity gain.\par

With the customized rank-$1$ channel \eqref{eq:H-DB-m}, $s$ is transmitted at the Tx with the precoder designed as follows:
\begin{equation}\label{eq-DB-F}
  {\bf f}_{\rm DB} = \sqrt{\frac{E}{K}}{\bf T}_{\rm DB}{\bf 1}_{K\times 1}.
\end{equation}
The signals arrive at the Rx for $M_{\rm R}$ reconfigurations are given by
\begin{equation}\label{eq:DB-r}
  {\bf{r}} = \left[{\bf{r}}_1,\,\ldots,\,{\bf{r}}_{M_{\rm R}}\right]^T = {\bf{H}}_{\rm DB}{\bf f}_{\rm DB}{ s} + {\bf{n}},
\end{equation}
where ${\bf{H}}_{\rm DB}=[{\bf{H}}_1,\,\ldots,\,{\bf{H}}_{M_{\rm R}}]^T$ and ${\bf{n}}=[{\bf{n}}_1,\,\ldots,\,{\bf{n}}_{M_{\rm R}}]^T$. The received signal is given as follows by employing the MRC combiner ${\bf w}_{\rm DB}={\bf H}{\bf f}_{\rm DB}/\| {{\bf{H}}{{\bf{f}}_{{\rm{DB}}}}} \|$:
\begin{equation}\label{eq:DB-y-1}
\begin{aligned}
  y  =& \left\| {{\bf{H}}_{{\rm{DB}}}{{\bf{f}}_{{\rm{DB}}}}} \right\|s + {\bf w}^H_{\rm DB}{\bf n}\\
  \approx & \sqrt {{\frac{E}{K}}\sum\limits_{m = 1}^{{M_{\rm{R}}}} {{{\bf{1}}_{1 \times K}}{\bf{\Xi }}_{{\rm{DB}},m}^H{\bf{R}}_{{\rm{DB}},m}^H{{\bf{R}}_{{\rm{DB}},m}}{{\bf{\Xi }}_{{\rm{DB}},m}}{{\bf{1}}_{K \times 1}}} } s \\
  &+ {\bf{w}}_{{\rm{DB}}}^H{\bf{n}}.
\end{aligned}
\end{equation}
We denote ${\bf R}_{{\rm DB},m}=[{\bf a}_{{\rm R},m,1},\,\ldots,\,{\bf a}_{{\rm R},m,K}]$ and rewrite \eqref{eq:DB-y-1} as follows for clarify of presentation:
\begin{equation}\label{eq:DB-y-2}
\begin{aligned}
  y \approx &\sqrt {\frac{E}{K}\sum\limits_{m = 1}^{{M_{\rm{R}}}} {\sum\limits_{k = 1}^K {\sum\limits_{n = 1}^K {\left[ {{{\bf{\Xi }}_{{\rm{DB}},m}}} \right]_{k,k}^*{{\left[ {{{\bf{\Xi }}_{{\rm{DB}},m}}} \right]}_{n,n}}{\bf{a}}_{{\rm{R}},m,k}^H{{\bf{a}}_{{\rm{R}},m,n}}} } } } s \\
  &+ {\bf{w}}_{{\rm{DB}}}^H{\bf{n}}\\
 \mathop  \approx \limits^{\left( c \right)}&  \sqrt {\frac{E}{K}\sum\limits_{m = 1}^{{M_{\rm{R}}}} {\sum\limits_{k = 1}^K {\sum\limits_{n = 1}^K {\left| {\left[ {{{\bf{\Xi }}_{{\rm{DB}},m}}} \right]_{k,k}^*{{\left[ {{{\bf{\Xi }}_{{\rm{DB}},m}}} \right]}_{n,n}}} \right|} } } } s + {\bf{w}}_{{\rm{DB}}}^H{\bf{n}},
\end{aligned}
\end{equation}
where ${( c )}$ is obtained because $|{\bf{a}}_{{\rm{R}},m,k}^H{{\bf{a}}_{{\rm{R}},m,n}}|\to 1$ and $\angle{[ {{{\bf{\Xi }}_{{\rm{DB}},m}}} ]_{k,k}^*{{[ {{{\bf{\Xi }}_{{\rm{DB}},m}}} ]}_{n,n}}{\bf{a}}_{{\rm{R}},m,k}^H{{\bf{a}}_{{\rm{R}},m,n}}}=0$ are guaranteed by the refining $\{{\bf \Gamma}_{k,m}\}_{m=1}^{M_{\rm R}}$ with common phase shifters similar to \eqref{eq:opt-RIS-com}. Given that ${\mathbb E}\{{\bf{w}}_{{\rm{DB}}}^H{\bf{n}}{\bf{n}}^H{\bf{w}}_{{\rm{DB}}}\}=\sigma^2$, the outage probability for the DB transmission is given as follows:
\begin{equation}\label{eq-OP-DB}
{{ P}_{{\rm{DB}}}} = \left\{\frac{{E {\sum\limits_{m = 1}^{{M_{\rm{R}}}} {\sum\limits_{k = 1}^K {\sum\limits_{n = 1}^K {{\left| {\left[ {{{\bf{\Xi }}_{{\rm{DB}},m}}} \right]_{k,k}^*{{\left[ {{{\bf{\Xi }}_{{\rm{DB}},m}}} \right]}_{n,n}}} \right|}} } } } }}{{K{\sigma ^2}}}<\gamma_{\rm th}\right\}.
\end{equation}
The outage probability for the BF transmission, ${{ P}_{{\rm{BF}}}}$, can be derived from ${{ P}_{{\rm{DB}}}}$ with $M_{\rm R}=1$. The SNR in \eqref{eq-OP-DB} implies that the DB transmission has a diversity order of $M_{\rm R}K^2$, which is $M_{\rm R}$-folds of the BF transmission\footnote{It is noted the BF transmission can be regarded as a diversity technique with diversity order of $K^2$.}. Therefore, increasing $M_{\rm R}$ can significantly reduce the outage probability and improve the transmission reliability. The SE for the DB transmission is approximately given by
\begin{equation}\label{eq:SE-DB}
\begin{aligned}
&{R_{{\rm{DB}}}}\\
&\approx \frac{1}{{{M_{\rm{R}}}}}{\log _2}\left( {1 + \frac{{E {\sum\limits_{m = 1}^{{M_{\rm{R}}}} {\sum\limits_{k = 1}^K {\sum\limits_{n = 1}^K {{\left| {\left[ {{{\bf{\Xi }}_{{\rm{DB}},m}}} \right]_{k,k}^*{{\left[ {{{\bf{\Xi }}_{{\rm{DB}},m}}} \right]}_{n,n}}} \right|}} } } } }}{{K{\sigma ^2}}}} \right)
\end{aligned}
\end{equation}
Applying Jensen's inequality, the ergodic SE for the DB transmission is upper bounded by
\begin{equation}\label{eq:SE-DB-upp}
\begin{aligned}
{{\bar R}_{{\rm{DB,upper}}}} = {\frac{1}{M_{\rm R}}}{\log _2}\left( 1 + \frac{{{M_{\rm R}}E{N_{\rm{T}}}{N_{\rm{R}}}\kappa }}{{{\sigma ^2}K{L_{\rm{R}}}\left( {\kappa  + 1} \right)}}\left( \sum\limits_{n = 1}^K {N_{{\rm{S,}}n}^2\rho _n^2} \right.\right.\\
\left.\left. + \frac{\pi }{4}\sum\limits_{n = 1,n \ne m}^K {\sum\limits_{m = 1}^K {{\rho _n}{\rho _m}{N_{{\rm{S,}}n}}{N_{{\rm{S,}}m}}} }  \right) \right).
\end{aligned}
\end{equation}
The DB transmission increases the SNR $M_{\rm R}$-folds while reducing the SE $M_{\rm R}$-folds compared with the BF transmission. Considering that $\frac{1}{M_{\rm R}}\log_2{(1+M_{\rm R}x)}\leq\log_2{(1+x)}$, ${{\bar R}_{{\rm{DB,upper}}}}\leq{{\bar R}_{{\rm{BF,upper}}}}$ always holds. Furthermore, the ergodic SE gap between the DB and the BF transmission continues to expand with an increasing $M_{\rm R}$ because $\log_2{(1+x)}-\frac{1}{M_{\rm R}}\log_2{(1+M_{\rm R}x)}$ is a monotone increasing function of $M_{\rm R}$. Thus, the DB transmission enhances the reliability with the diversity order of $M_{\rm R}K^2$ at the cost of SE decline.\par

Although we focus on the downlink in this study, the proposed transmission schemes can be extended to the uplink. Considering that the reliability can be enhanced with a low-complexity Tx (designed with statistical CSI and without increasing the symbol transmission rate), the DS and DB transmissions are particularly suited for uplink because the user equipment can be easily designed and the multi-reception with a higher ADC sampling rate will be solved by the powerful base station.
\section{Numerical Results}\label{sec:5}

\begin{figure}[!t]
\centering
\includegraphics[width=0.5\textwidth]{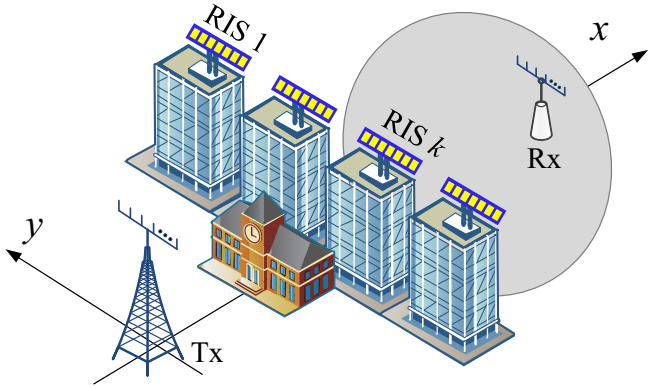}
\caption{Deployment of the considered multi-RISs assisted MIMO system.}
\label{Fig.deployment} 
\end{figure}

In this section, we provide numerical results to illustrate the performance of the proposed multi-timescale channel customization for transmission design in the RIS-assisted MIMO systems. The throughput and reliability of the transmission schemes with the customized channel are evaluated in a sub-$6$ GHz\footnote{ {Note that this work is not restricted to  sub-$6$ GHz systems, but can be applied to any wireless systems where EM waves propagate through multipath.}} MIMO system assisted by multiple RISs. The system carrier frequency is $f_c=3.5$ GHz. Unless otherwise specified, the Tx has $N_{\rm T}=16$ transmit antennas, and the Rx has $N_{\rm R}=4$ receive antennas. In Fig. \ref{Fig.deployment}, the Tx is deployed in the original Cartesian coordinates, and the Rx is randomly located in a circular blind coverage with the center being $[200,0]^T$ and the radius being $50$. Four RISs are mounted on the DFT directions of the Tx with coordinate $x=150$ to provide reliable transmission links between the Tx and the Rx. To combat different path losses, we assume that the number of elements in each RIS is inversely proportional to the associated path loss, that is, ${N_{{\rm{S,}}k}\rho _k}=C$, where $C$ determines $N_{{\rm{S,}}k}$. Under the considered deployment, we initialize $C=10^{-6}$ and have $\{N_{{\rm{S,}}1},N_{{\rm{S,}}2},N_{{\rm{S,}}3},N_{{\rm{S,}}4}\}=\{211,174,174,211\}$. Considering that the Tx and RISs are mounted at buildings, the Rician factor and the number of NLoS paths for the Tx--RISs channel are set to $\kappa=10$ dB and $L_{\rm T}=2$, respectively. The number of paths between the RIS--Rx channel is set to $L_{\rm R}=10$ due to the abundant scatters around the Rx. The AoD and AoA of the NLoS paths are calculated via the positions of the Tx, Rx, RISs, and the random scatters. The noise power is set to $-100$ dBm.

\subsection{Performance Evaluation for the SM and BF Transmissions}
In this subsection, we examine the SE performance of the SM and BF transmissions with the customized channel. The tightness of the upper bound and approximation of the ergodic SE is displayed under different factors. The crossing-point on the upper bounds of the SM and BF transmissions is demonstrated to show the relative SE performance.\par

\begin{figure}[!t]
\centering
\subfigure[The SM transmission]{
\label{Fig.SE-SM-E} 
\includegraphics[width=0.5\textwidth]{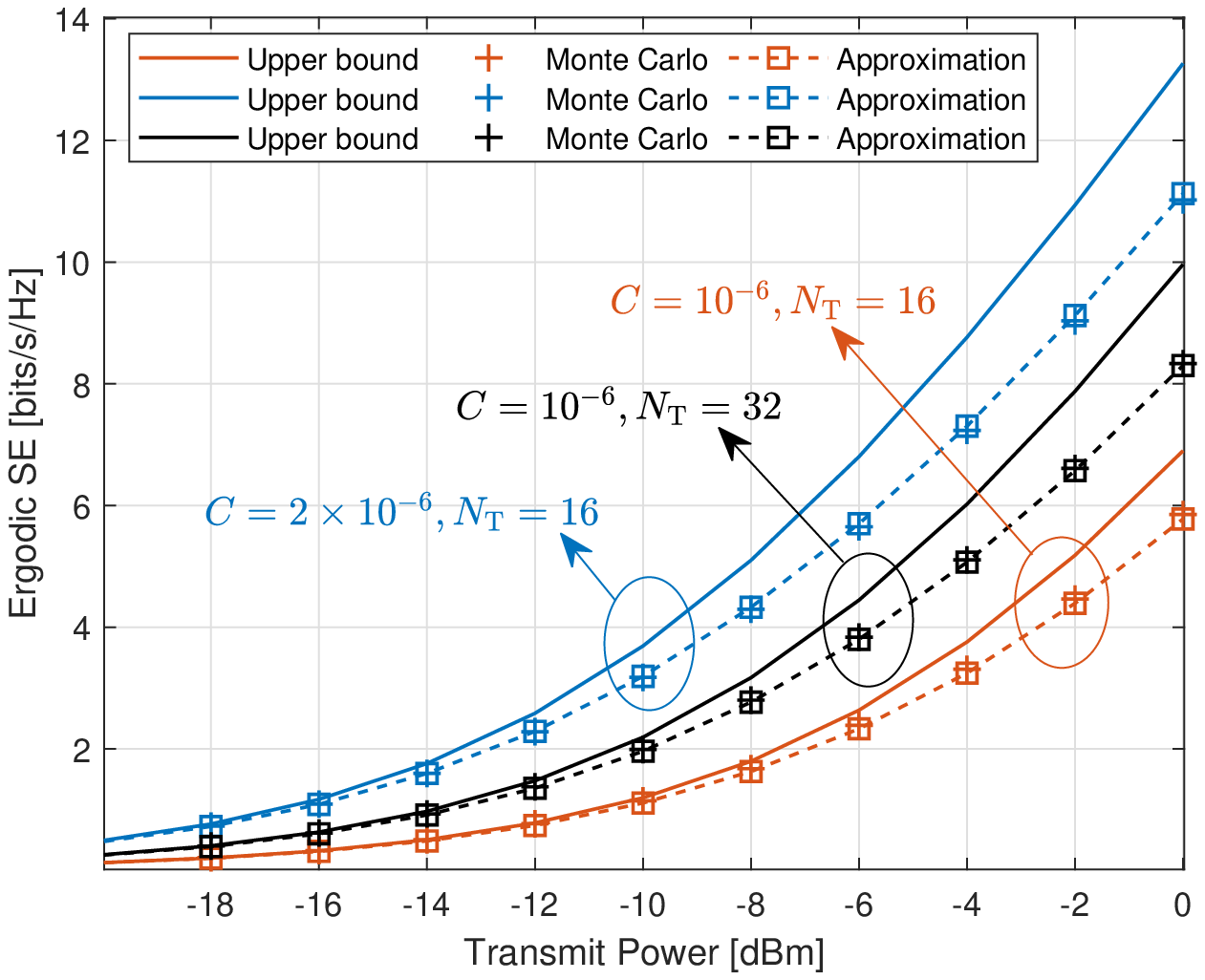}}
\subfigure[The BF transmission]{
\label{Fig.SE-BF-E} 
\includegraphics[width=0.5\textwidth]{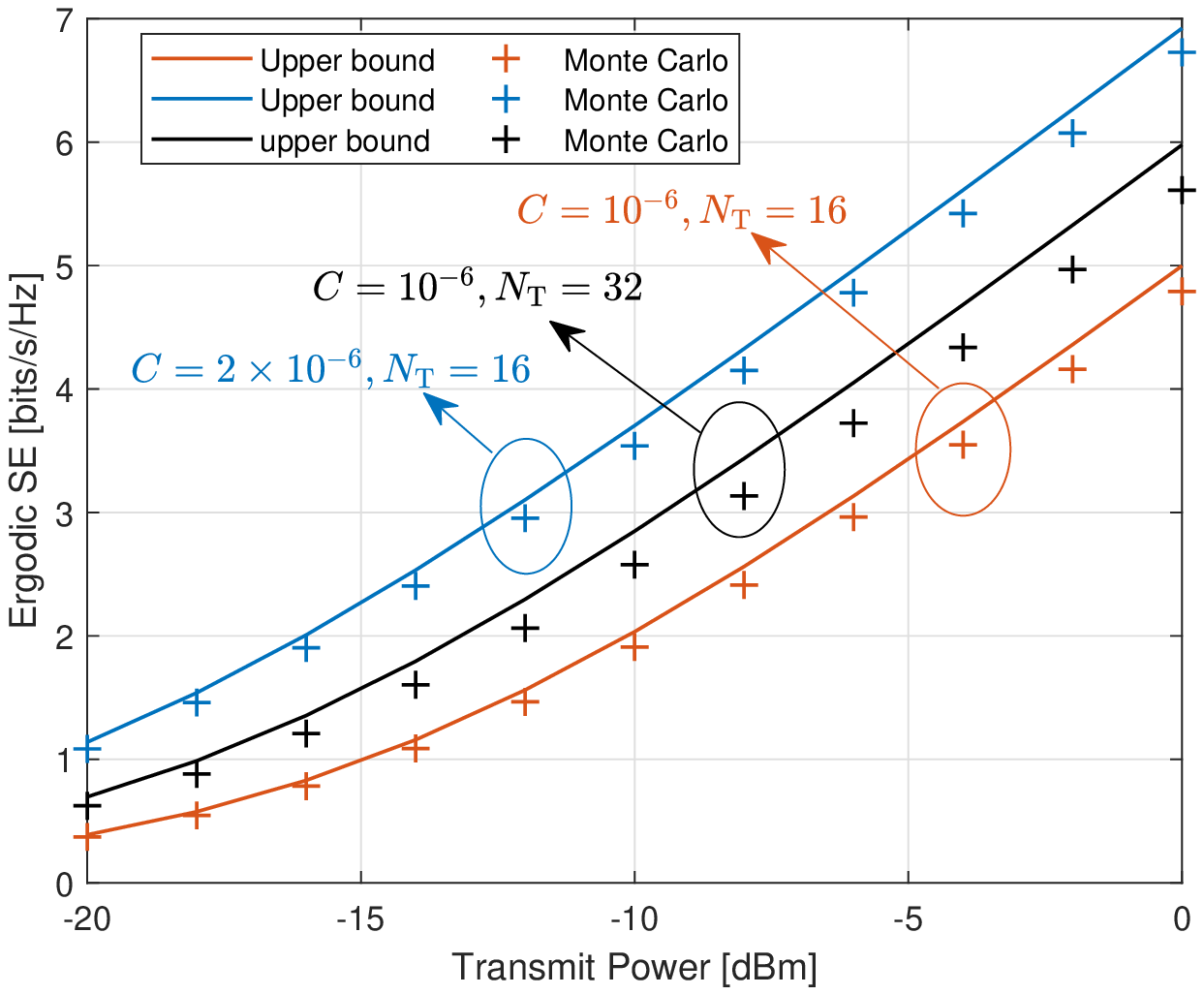}}
\caption{Ergodic SE versus transmit power with different system setups.}
\label{Fig.SE-E} 
\end{figure}

Fig. \ref{Fig.SE-E} presents the ergodic SE of the SM and BF transmissions versus the transmit power with different system setups. In Fig. \ref{Fig.SE-SM-E} for the SM transmission, the closed-form approximations of the ergodic SE match well with the Monte Carlo results, while the upper bounds are loose at the high ergodic SE regime. Although the closed-form expression for the ergodic SE of the BF are difficult to obtain, the derived upper bound is tight enough to be an approximation, as shown in Fig. \ref{Fig.SE-BF-E}.  {The ergodic SE for the SM and BF transmissions increase with the array size at the Tx and RIS, that is, $N_{\rm T}$ and $N_{{\rm S},k}$ ($N_{{\rm S},k}$ is embodied by $C$). A larger ergodic SE gain can be obtained by a doubled $N_{{\rm S},k}$ than a doubled $N_{\rm T}$, which can be explained through \emph{Theorems \ref{Them:SM-upp}} and \emph{ \ref{Them:BF-upp}} that $N_{{\rm S},k}$ quadratically increases the SNR, while $N_{\rm T}$ can only bring a linear improvement to the SNR.}\par

\begin{figure}[!t]
\centering
\subfigure[The SM transmission]{
\label{Fig.SE-SM-k} 
\includegraphics[width=0.5\textwidth]{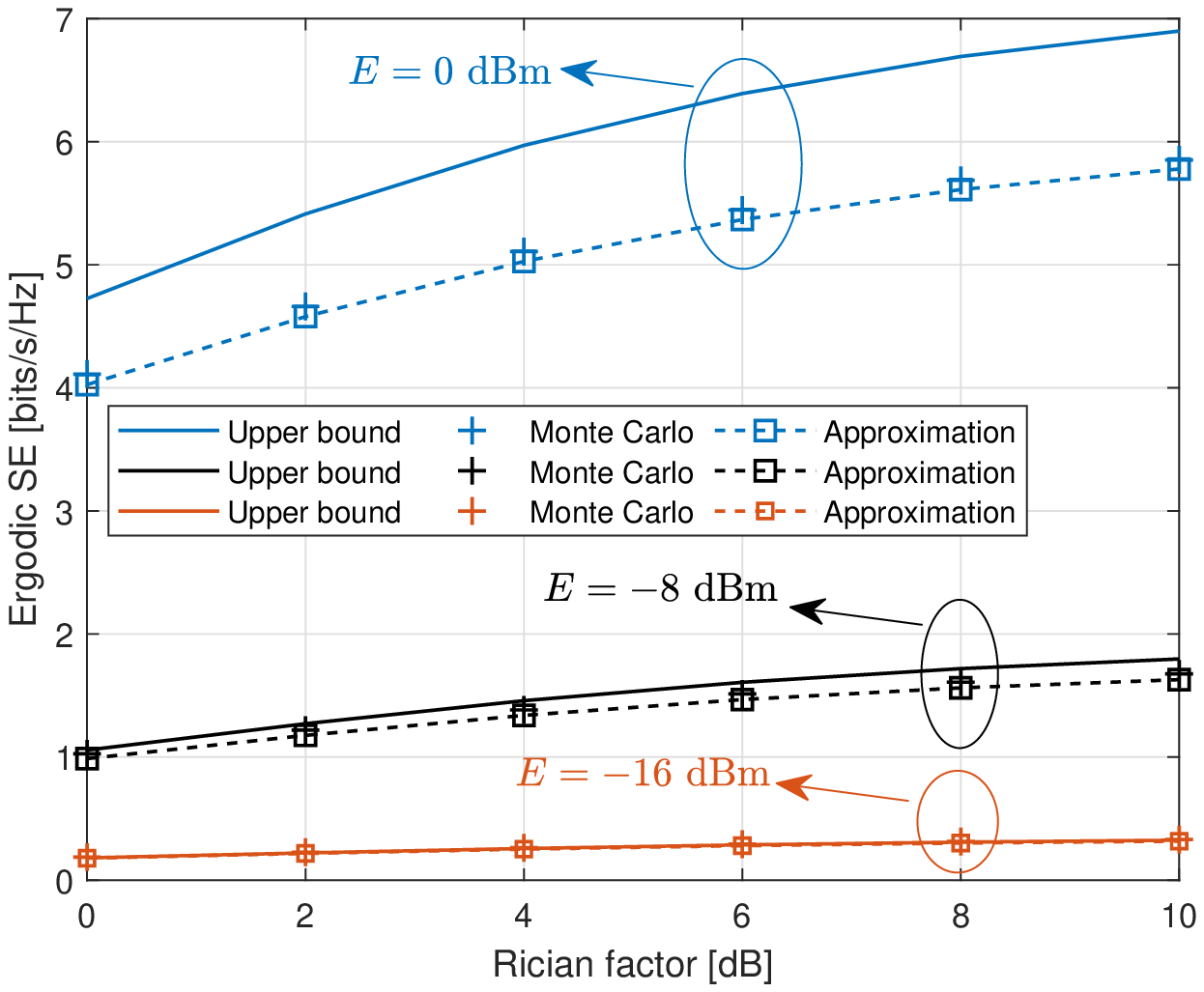}}
\subfigure[The BF transmission]{
\label{Fig.SE-BF-k} 
\includegraphics[width=0.5\textwidth]{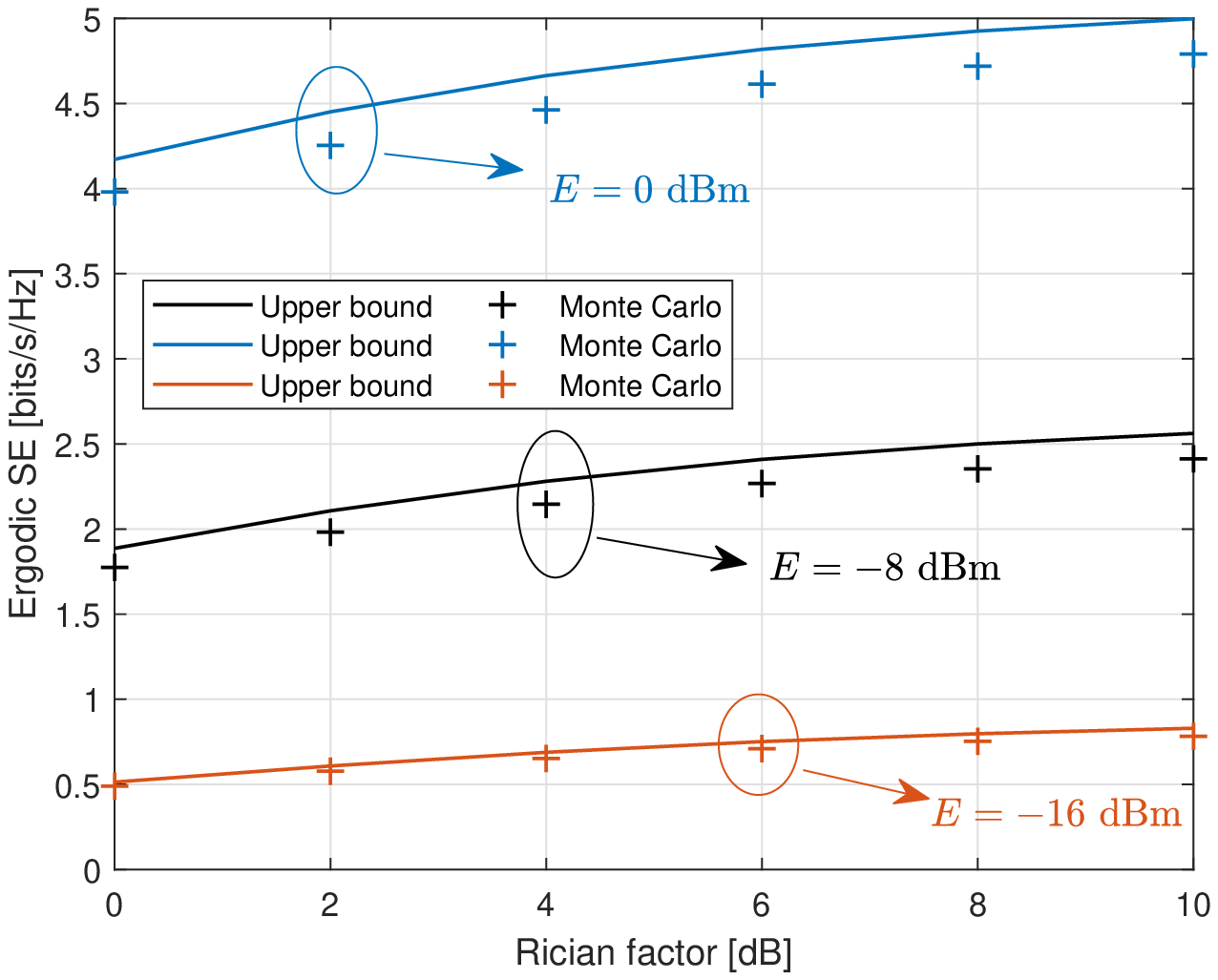}}
\caption{Ergodic SE versus the Rician factor at the TX--RISs channel.}
\label{Fig.SE-k} 
\end{figure}

The ergodic SE of the SM and BF transmissions versus the Rician factor at the TX--RISs channel is shown in Fig. \ref{Fig.SE-k}. As discussed in \emph{Theorems \ref{Them:SM-upp}} and \emph{\ref{Them:BF-upp}}, a larger Rician factor improves the ergodic SE. The reason for this improvement is that the energy leakage is less when the LoS path, which is selected as the transmission path, gradually nominating the Tx--RISs channel. In the SM transmission, the gaps between the upper bound and the Monte Carlo result are enlarged by increasing the Rician factor. However, these gaps remain stable in the BF transmission, indicating that the upper bound well depicts the ergodic SE of the BF transmission.\par

\begin{figure}[!t]
\centering
\subfigure[The SM transmission]{
\label{Fig.SE-SM-L} 
\includegraphics[width=0.5\textwidth]{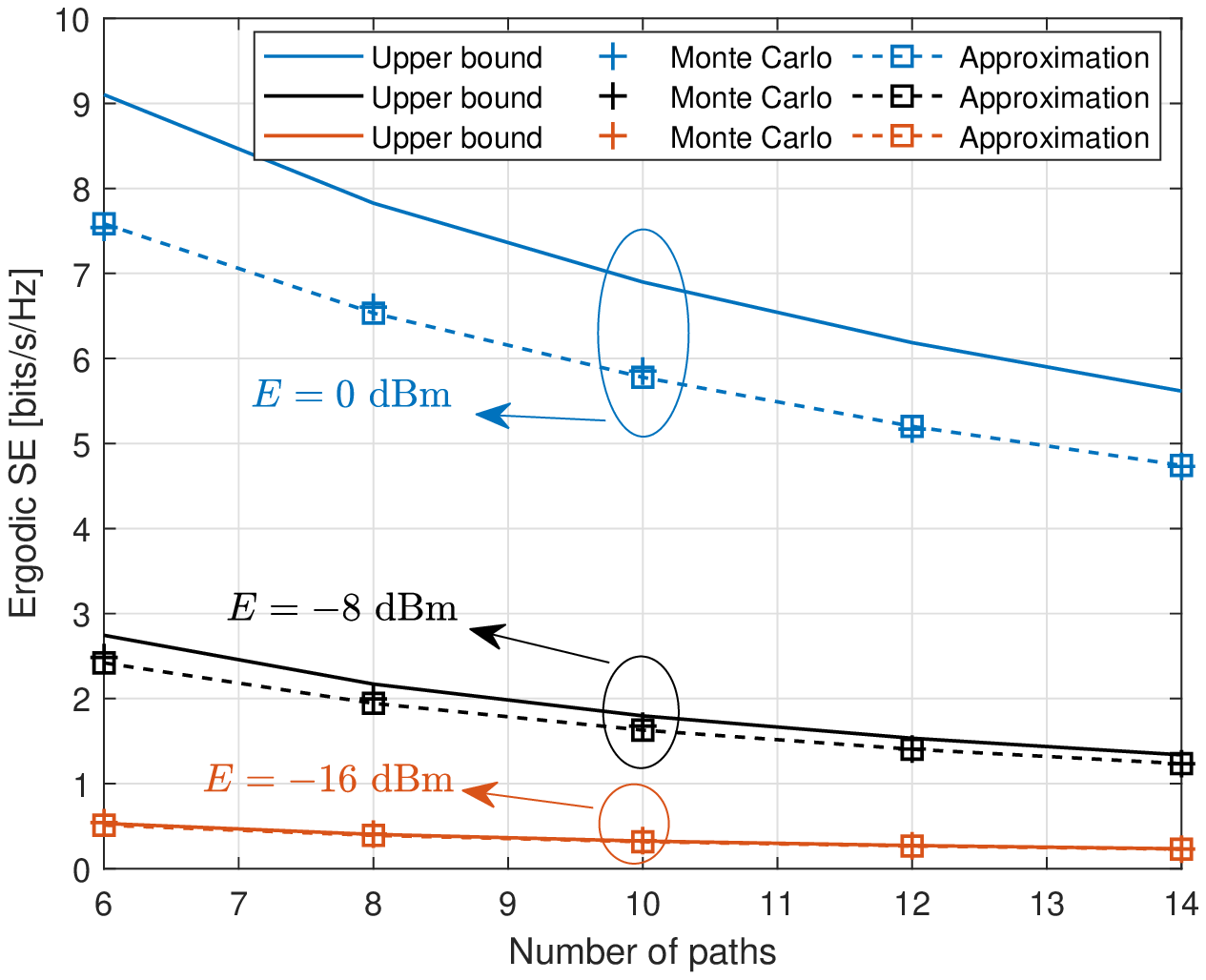}}
\subfigure[The BF transmission]{
\label{Fig.SE-BF-L} 
\includegraphics[width=0.5\textwidth]{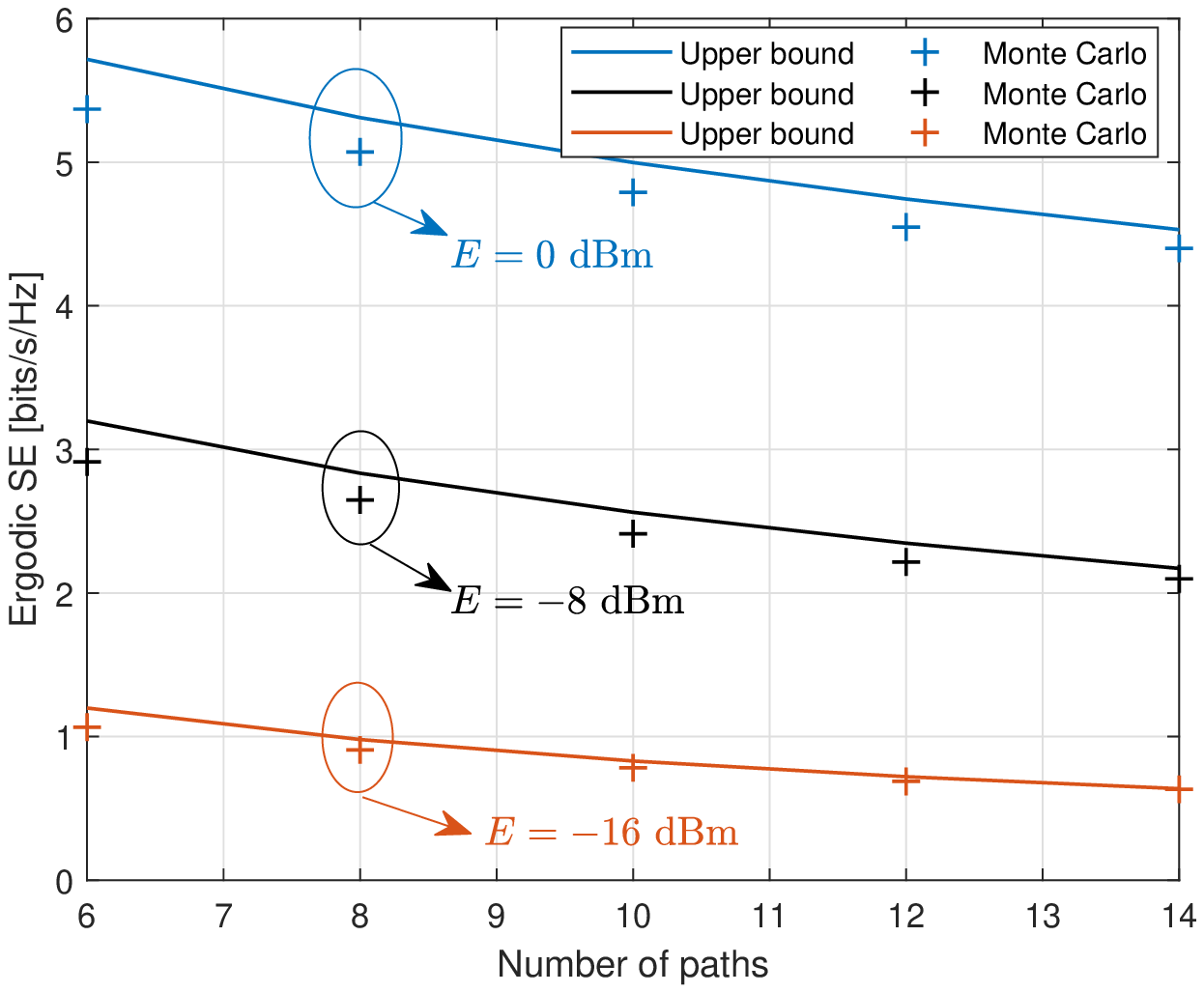}}
\caption{Ergodic SE versus the number of paths at the RISs-Rx channel.}
\label{Fig.SE-L} 
\end{figure}

In Fig. \ref{Fig.SE-L}, the ergodic SE versus the number of paths at the RISs--Rx channel is demonstrated. The increasing number of paths reduces the ergodic SE, exhibiting an opposite trend as compared with Fig. \ref{Fig.SE-k}. These opposite results coincide with the analytical expression in \emph{Theorems \ref{Them:SM-upp}} and \emph{ \ref{Them:BF-upp}} and have the same principle. When the inactive paths are fewer and weaker, the energy leakage decreases and the ergodic SE can be improved.\par

\begin{figure}[!t]
\centering
\subfigure[The SM transmission]{
\label{Fig.SE-SM-CE} 
\includegraphics[width=0.5\textwidth]{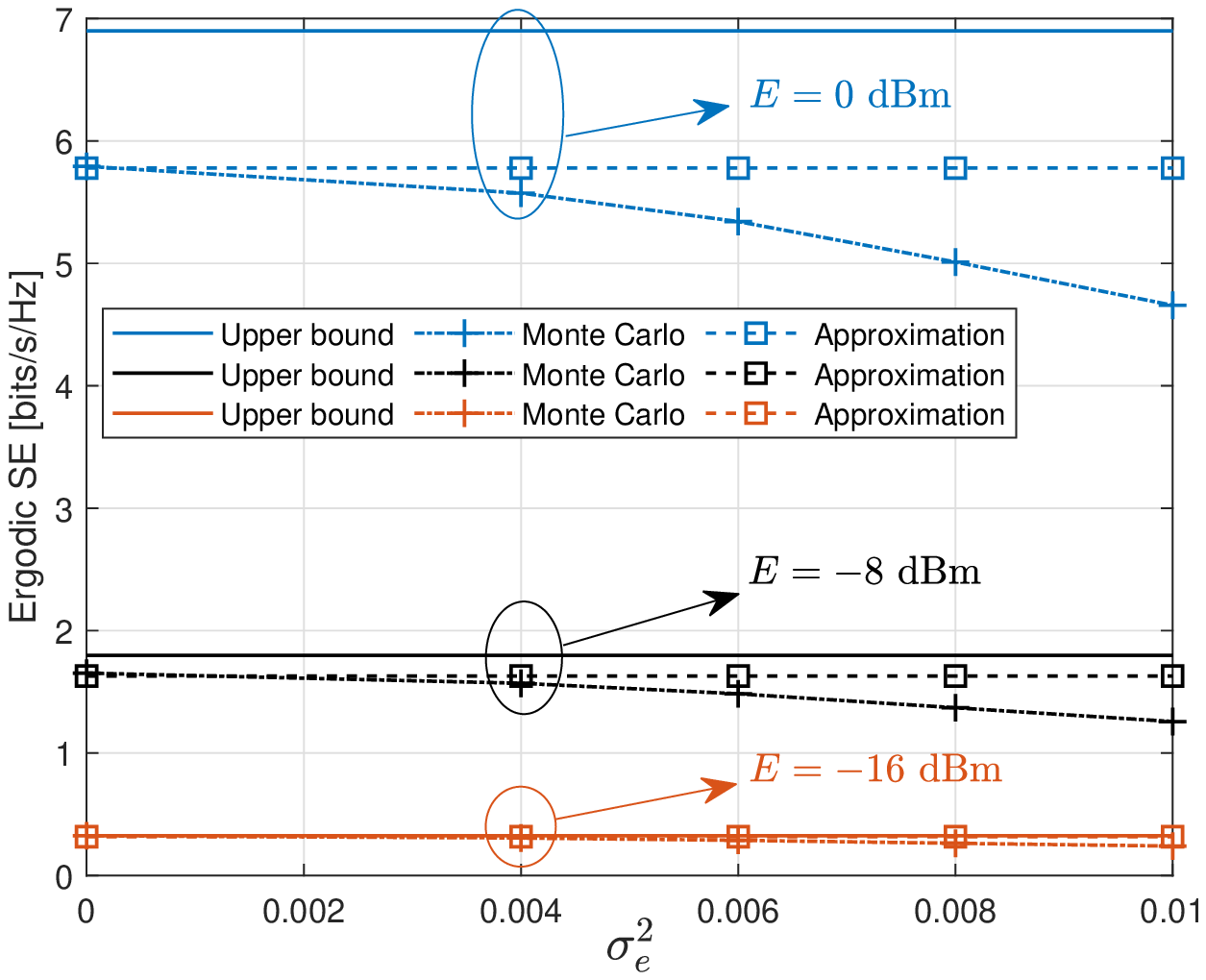}}
\subfigure[The BF transmission]{
\label{Fig.SE-BF-CE} 
\includegraphics[width=0.5\textwidth]{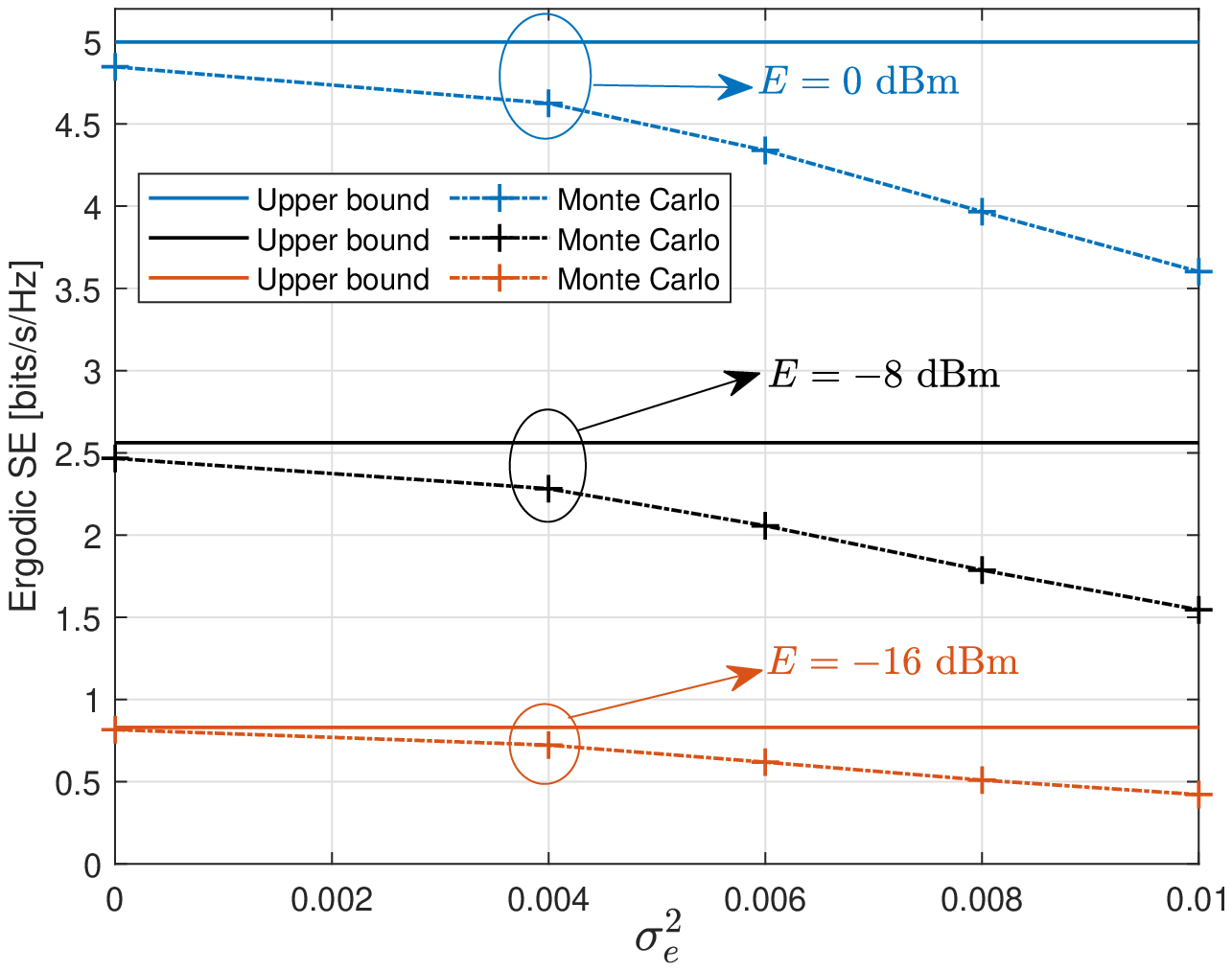}}
\caption{Ergodic SE versus channel estimation error with different transmit power.}
\label{Fig.SE-CE} 
\end{figure}

In the proposed multi-timescale channel customization for transmission design, the AoA and AoD are the most crucial channel parameters. Accordingly, we study the impacts of the angle estimation error on the ergodic SE for the SM and BF transmission, as shown in Fig. \ref{Fig.SE-CE}. Considering that the LoS paths selected from the Tx--RISs channel are quasi-static, the AoA and AoD of these paths can be precisely obtained. Consequently, we focus on the estimation error of the AoA and AoD in the RISs--Rx channel. The AoA/AoD at the RISs--Rx channel with estimation error is given by ${{\hat\Phi} _{{k},l}^{\rm{A/D}}}={{\Phi} _{{k},l}^{\rm{A/D}}}+n_e$, where $n_e \in {\mathcal {CN}}(0,\sigma_e^2)$. The upper bounds and approximation are used as baselines because they are derived with perfect parameters. The Monte Carlo results show that the ergodic SE decreases when the estimation error increases. Furthermore, the performance reduction is enlarged with an increasing transmit power,  because the energy leakage brought by the imperfect CSI is larger.\par

\begin{figure}[!t]
\centering
\includegraphics[width=0.5\textwidth]{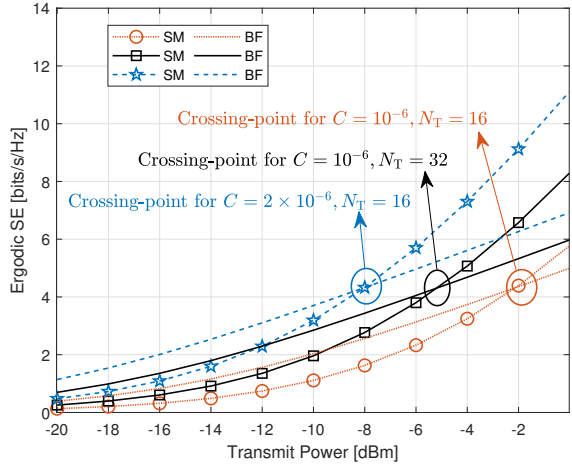}
\caption{Relative ergodic SE performance for the SM and BF transmissions.}
\label{Fig.CP} 
\end{figure}

The relative ergodic SE performance for the SM and BF transmissions is illustrated in Fig. \ref{Fig.CP}. When the array size at the Tx and RIS increases, the crossing-point for the ergodic SE of the SM and BF transmissions shifts to the low transmit power regime. This result reveals that the superiority of the SM transmission can be expanded by increasing the number of antennas and the number of RIS elements, which is consistent with the analysis in Section \ref{sec:CP}. Moreover, the transmit power threshold reduces by $3$ and  $6$ dB when the $N_{\rm T}$ and $C$ are doubled, respectively. This result is the same as the analysis for the special case with $N_{\rm R}=2$.

\subsection{Tradeoff between the Throughput and Reliability}

In this subsection, we evaluate the tradeoff between the throughput and reliability of the proposed channel customization within symbol times via the ergodic SE and BER performance. When $M_{\rm R}=1$, the time diversity within symbol times disappears, and the DS/DB transmission is reduced to the SM/BF transmission.\par

\begin{figure}[!t]
\centering
\includegraphics[width=0.5\textwidth]{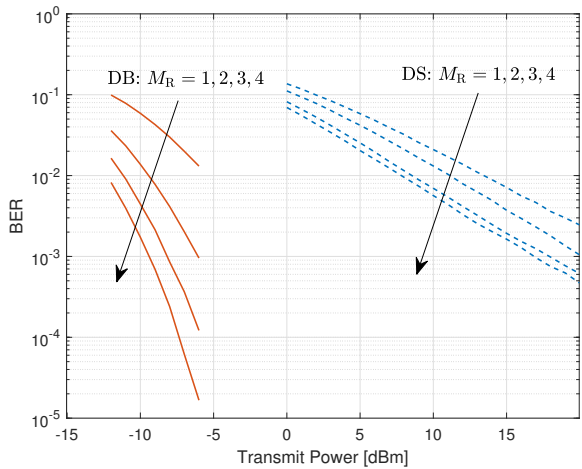}
\caption{Comparison of the BER for the different transmissions.}
\label{Fig.BER} 
\end{figure}

\begin{figure}[!t]
\centering
\subfigure[The DS transmission]{
\label{Fig.SE-DS} 
\includegraphics[width=0.5\textwidth]{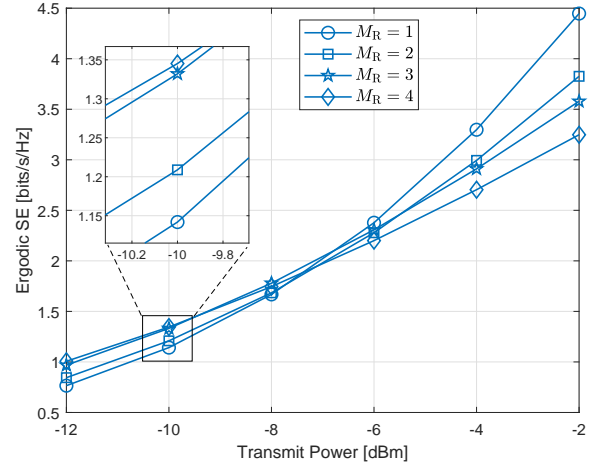}}
\subfigure[The DB transmission]{
\label{Fig.SE-DB} 
\includegraphics[width=0.5\textwidth]{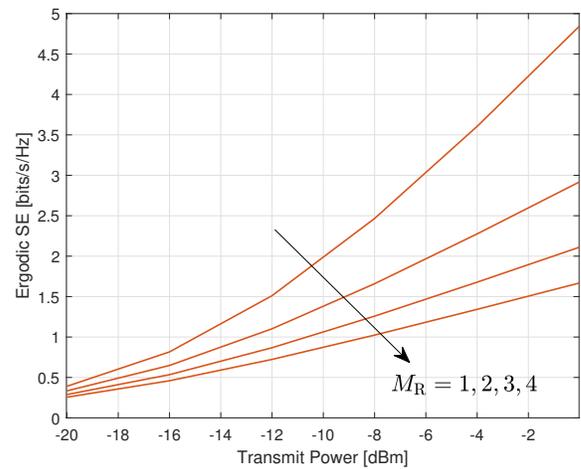}}
\caption{Ergodic SE versus transmit power with different numbers of RIS reconfiguration within symbol times.}
\label{Fig.SE-DS-DB} 
\end{figure}

In Fig. \ref{Fig.BER}, we compare the BER for the SM (DS with $M_{\rm R}=1$), BF (DB with $M_{\rm R}=1$), DS, and DB transmission. The SM transmission has the worst BER performance while the DB with a larger $M_{\rm R}$ gets the best BER performance even though its ergodic SE is the lowest, as shown in Fig. \ref{Fig.SE-DB}. The BER of the DB transmission is far better than that of the DS transmission because the diversity order of the DB and DS transmissions are $M_{\rm R}K^2$ and $M_{\rm R}$, respectively. Increasing $M_{\rm R}$ to obtain a larger diversity order can significantly enhance the reliability. We consider the special cases where the DB and DS transmit single-stream and full-stream data, respectively. When the number of data stream is set between $[1,N_{\rm R}]$, the BER performance will lie in the transitional zone between the DB and the DS transmissions.\par

Fig. \ref{Fig.SE-DS-DB} shows the ergodic SE for the SM (DS with $M_{\rm R}=1$), BF (DB with $M_{\rm R}=1$), DS, and DB transmission. In Fig. \ref{Fig.SE-DS}, the DS transmission has opposite SE performance in the low and high transmit power regimes. In the low transmit power regime where the noise power determines the SE, increasing $M_{\rm R}$ can bring diversity and coherent gain to increase the SNR, as detailed at the end of Section \ref{sec:DS}. Accordingly, the SE performance is improved. Considering that the required bandwidth or retransmission symbols increases with $M_{\rm R}$, the ergodic SE decreases in the high transmit power regime. The DB transmission increases the SNR $M_{\rm R}$-folds while reducing the SE $M_{\rm R}$-folds, as detailed in Section \ref{sec:DB}. Thus, Fig. \ref{Fig.SE-DB} shows that the ergodic SE of the DB transmission with $M_{\rm R}>1$ is always below the BF transmission ($M_{\rm R}=1$). Moreover, the gap between the DB and the BF transmission increases with $M_{\rm R}$.\par
The comparison of Fig. \ref{Fig.SE-DS-DB} with Fig. \ref{Fig.BER} demonstrated that the sacrifice of throughput is paid with the enhanced reliability. Therefore, the tradeoff between the throughput and the reliability can be realized by properly choosing the number of data streams and $M_{\rm R}$.\par


\section{Conclusion}\label{sec:6}
 {In this study, we customized the composite channel between the Tx and the Rx by reconfiguring the phase shifters of RISs in RIS-assisted systems to facilitate transmission designs, given that MIMO transmission techniques depend on specific channel characteristics.} The proposed multi-timescale channel customization sheds light on how channel characteristics promote the performance of transmission schemes.  {To improve the SE, we utilized the statistical CSI in angle-coherent times to customize a high-rank channel for SM transmission in the interference-limited regime. In the noise-limited regime, where the BF transmission is adopted, we customized a strongly correlated channel with rank-1 by tuning the phase shifters of the RIS. To enhance the system's reliability by achieving diversity, we proposed DS and DB transmissions based on SM and BF transmissions, with channel customization executed within symbol times.} Numerical results demonstrated that the tradeoff between throughput and reliability of the MIMO system can be achieved by flexibly customizing the composite channel for transmission designs.\par

\appendices
\section{}\label{App:A}
Since $\beta_{n,{\rm R},l_n}\sim \mathcal{CN}( 0,1)$, $x=|\sqrt{2}\beta_{n,{\rm R},l_n}|^2$ is a chi-squared distributed random variable with $2$ degrees of freedom. Therefore, the  {probability density function (PDF)} of $x$ is given by
\begin{equation}\label{eq:App-SM-4}
  f\left( x \right) = \left\{ {\begin{aligned}
&{\frac{1}{2}{e^{ - \frac{1}{2}x}}},x> 0\\
&{0,x \le 0}
\end{aligned}} \right..
\end{equation}
Given the  {PDF} of $x$, we have
\begin{equation}\label{eq:App-SM-5}
  \begin{aligned}
{{\bar R}_{{\rm{SM}},n}} &= \int\nolimits_0^\infty  {{{{\log }_2}\left( {1 + \frac{x}{2c_n}} \right)}f\left( x \right)dx}  \\
&= \frac{1}{{2\ln 2}}\int\nolimits_0^\infty  {\ln \left( {1 + \frac{x}{2c_n}} \right){e^{ - \frac{1}{2}x}}dx} \\
&\mathop  = \limits^{\left( a \right)} \frac{1}{{2\ln 2}}\left(-2 {{e^{c_n}}{\rm{Ei}}\left( { - c_n} \right)} \right)  \\
&= -\frac{1}{{\ln 2}} {{e^{c_n}}{\rm{Ei}}\left( { - c_n} \right)},
\end{aligned}
\end{equation}
where $(a)$ uses the property \cite{Table}
\begin{equation}\label{eq:App-SM-6}
  \int\nolimits_0^\infty  {{e^{ - \mu x}}\ln \left( {1 + \beta x} \right)dx} = -{\frac{1}{\mu}}e^{\frac{\mu}{\beta}}{\rm Ei}\left(-{\frac{\mu}{\beta}}\right).
\end{equation}
Substituting \eqref{eq:App-SM-5} into \eqref{eq:App-SM-1} completes the proof.

\section{}\label{App:C}
Applying Jensen's inequality, the ergodic SE is upper bounded by
\begin{equation}\label{eq:BF-upp-e1}
  \begin{aligned}
{{\bar R}_{{\rm{BF}}}} &\approx {\mathbb E}\left\{ {{{\log }_2}\left( {1 + \frac{{E\sum\limits_{n = 1}^K {\sum\limits_{m = 1}^K {\left| {\left[ {{{\bf{\Xi }}_{{\rm{BF}}}}} \right]_{n,n}^*{{\left[ {{{\bf{\Xi }}_{{\rm{BF}}}}} \right]}_{m,m}}} \right|} } }}{{{\sigma ^2}K}}} \right)} \right\}\\
 &\le {\log _2}\left( {1 + \frac{{E\sum\limits_{n = 1}^K {\sum\limits_{m = 1}^K {\mathbb E}{\left\{ {\left| {\left[ {{{\bf{\Xi }}_{{\rm{BF}}}}} \right]_{n,n}^*{{\left[ {{{\bf{\Xi }}_{{\rm{BF}}}}} \right]}_{m,m}}} \right|} \right\}} } }}{{{\sigma ^2}K}}} \right).
\end{aligned}
\end{equation}
Expanding the elements in ${{{\bf{\Xi }}_{{\rm{BF}}}}}$, we have
\begin{equation}\label{eq:BF-upp-e2}
  \begin{aligned}
&{\mathbb E}\left\{ {\left| {\left[ {{{\bf{\Xi }}_{{\rm{BF}}}}} \right]_{n,n}^*{{\left[ {{{\bf{\Xi }}_{{\rm{BF}}}}} \right]}_{m,m}}} \right|} \right\} \\
= &{\mathbb E}\left\{ {\left| {{\rho _n}\alpha _{n,{\rm{R}},{l_n}}^*\alpha _{{\rm{T}},n,0}^*{\rho _m}{\alpha _{m,{\rm{R}},{l_m}}}{\alpha _{{\rm{T}},m,0}}} \right|} \right\}\\
 =  &\frac{{{N_{\rm{T}}}{N_{\rm{R}}}\kappa }}{{{L_{\rm{R}}}\left( {\kappa  + 1} \right)}}{N_{{\rm{S}},n}}{N_{{\rm{S}},m}}{\rho _n}{\rho _m}{\mathbb E}\left\{ {\left| {\beta _{n,{\rm{R}},{l_n}}^*{\beta _{m,{\rm{R}},{l_m}}}} \right|} \right\}.
\end{aligned}
\end{equation}
When $m=n$, \eqref{eq:BF-upp-e2} can be deduced as
\begin{equation}\label{eq:BF-upp-e3}
  \begin{aligned}
{\mathbb E}\left\{ {\left| {\left[ {{{\bf{\Xi }}_{{\rm{BF}}}}} \right]_{n,n}} \right|}^2 \right\} &=  \frac{{{N_{\rm{T}}}{N_{\rm{R}}}\kappa }}{{{L_{\rm{R}}}\left( {\kappa  + 1} \right)}}{N^2_{{\rm{S}},n}}{\rho^2 _n}{\mathbb E}\left\{ {\left| {\beta _{n,{\rm{R}},{l_n}}} \right|^2} \right\}\\
&=\frac{{{N_{\rm{T}}}{N_{\rm{R}}}\kappa }}{{{L_{\rm{R}}}\left( {\kappa  + 1} \right)}}{N^2_{{\rm{S}},n}}{\rho^2 _n}.
\end{aligned}
\end{equation}
because $\beta _{n,{\rm{R}},{l_n}}\in {\mathcal {CN}}(0,1)$. When $m \ne n $, the expectation in \eqref{eq:BF-upp-e2} can be rewritten as
\begin{equation}\label{eq:BF-upp-e4}
  \begin{aligned}
  &{\mathbb E}\left\{ {\left| {\beta _{n,{\rm{R}},{l_n}}^*{\beta _{m,{\rm{R}},{l_m}}}} \right|} \right\} \\
  = &\frac{1}{2}{\mathbb E}\left\{ \sqrt{ {\left|\sqrt{2} {\beta _{n,{\rm{R}},{l_n}}} \right|^2}} \right\}   {\mathbb E}\left\{ \sqrt{{\left|\sqrt{2} {{\beta _{m,{\rm{R}},{l_m}}}} \right|^2}} \right\}.
\end{aligned}
\end{equation}
It is clarified in Appendix \ref{App:A} that
$x=|\sqrt{2}\beta_{n,{\rm R},l_n}|^2$ is a chi-squared distributed random variable with $2$ degrees of freedom. Therefore, we have
\begin{equation}\label{eq:BF-upp-e5}
\begin{aligned}
  &{\mathbb E}\left\{ \sqrt{ {\left|\sqrt{2} {\beta _{n,{\rm{R}},{l_n}}} \right|^2}} \right\}\\
  =&{\mathbb E}\left\{ \sqrt{ x} \right\}= \frac{1}{2}\int_0^\infty  {\sqrt x {e^{ - \frac{x}{2}}}dx}\\
   \mathop  = \limits^{(b)}  & \frac{1}{2}\sqrt \pi  {2^{ - 1}}{\left( {\frac{1}{2}} \right)^{ - 1 - \frac{1}{2}}}\left( {2 - 1} \right)!!= \sqrt {\frac{\pi }{2}},
\end{aligned}
\end{equation}
where $(b)$ utilizes the property \cite{Table}
\begin{equation}\label{eq:BF-upp-e6}
 \int_0^\infty  { x^{n-\frac{1}{2}} {e^{ -\mu x}}dx}=\sqrt{\pi}2^{-n}\mu^{-n-\frac{1}{2}}(2n-1)!! ,
\end{equation}
with $n=1$ and $\mu=\frac{1}{2}$. Substituting \eqref{eq:BF-upp-e5} into \eqref{eq:BF-upp-e4}, we have
\begin{equation}\label{eq:BF-upp-e7}
  {\mathbb E}\left\{ {\left| {\beta _{n,{\rm{R}},{l_n}}^*{\beta _{m,{\rm{R}},{l_m}}}} \right|} \right\} = \frac{\pi}{4}.
\end{equation}
When $m\ne n$, \eqref{eq:BF-upp-e2} can be derived as
\begin{equation}\label{eq:BF-upp-e8}
{\mathbb E}\left\{ {\left| {\left[ {{{\bf{\Xi }}_{{\rm{BF}}}}} \right]_{n,n}^*{{\left[ {{{\bf{\Xi }}_{{\rm{BF}}}}} \right]}_{m,m}}} \right|} \right\} =   \frac{\pi}{4}\frac{{{N_{\rm{T}}}{N_{\rm{R}}}\kappa }}{{{L_{\rm{R}}}\left( {\kappa  + 1} \right)}}{N_{{\rm{S}},n}}{N_{{\rm{S}},m}}{\rho _n}{\rho _m}.
\end{equation}
Substituting \eqref{eq:BF-upp-e3} and \eqref{eq:BF-upp-e8} into \eqref{eq:BF-upp-e1}, the proof in completed.

\begin{small}

\end{small}
%

\end{document}